\documentclass[12pt,preprint]{aastex}
\usepackage{url}
\usepackage{longtable}
\usepackage{emulateapj5}

\begin{document}
\title{Color gradients in galaxies out to $z\sim3$: dependence
  on galaxy properties
}

\author{Niraj Welikala\altaffilmark{1}, Jean-Paul Kneib\altaffilmark{2}  
       }

\altaffiltext{1}{Insitut d'Astrophysique Spatiale, B\^atiment 121, Universit\'e Paris-Sud XI \& CNRS, 91405 Orsay Cedex, France, niraj.welikala@ias.u-psud.fr}
\altaffiltext{2}{Laboratoire d'Astrophysique de Marseille, CNRS \&
  Universit\'e Aix-Marseille, 38, rue Fr\'ed\'eric Joliot-Curie, 13388
  Marseille Cedex 13, France, jean-paul.kneib@oamp.fr}

\begin{abstract}
\label{sec:abstract}

Using HST/ACS observations, we measure the color gradients of 3248
galaxies in the GOODS-South field out to $z\sim3$ and $i_{AB}<25.5$ 
and characterize their dependence on galaxy properties (luminosity,
apparent magnitude, galaxy size, redshift and morphological type). The
color gradient is measured by the difference of $v-i$ color outside ($R_{50}\leq r<2R_{50}$)
and inside the half light radius. The gradient shows little evolution
with redshift up to $z\sim1$ but increases from $z\sim1$ to $z\sim2$ before
flattening out. It also increases with apparent magnitude, with a
median value of 0.24 magnitudes at $i_{AB}\sim25.5$. It has a strong
color dependence, with the bluest galaxies (in terms of observed
color) having cores that are bluer relative to their
outskirts. We probe the redshift evolution by stacking galaxies and measuring the radial variation
of $v-i$ within them. At low redshifts ($z<0.5$), the centres of galaxies ($r<R_{50}$) are slightly
redder than their outskirts ($1.5R_{50}\leq r<2R_{50}$). 
Galaxies at $z\sim1$ and  $-22.0<M_{I}\le -21.0$ are bluer in
their cores by 0.1 magnitudes, on average, compared to their outskirts. 
For $z>1$, galaxies show increasingly bluer cores while the color of the
outskirts does not change as rapidly. At $z\sim2.5$ and $-22.0<M_{I}\le -21.0$, we observe a difference, on average, of 0.4 magnitudes between
the centre and the outskirts. The observed color gradients may indicate that strong star
formation in galaxies at $z\gtrsim2$ is concentrated in their central
regions. These color gradients and their dependence on observable
properties could also have a significant impact on shear measurements in upcoming weak lensing cosmological surveys.

\end{abstract}

\keywords{galaxies:evolution --- galaxies:statistics --- cosmology:systematics}

\section{INTRODUCTION
\label{sec:intro}}

In order to provide useful constraints on dark energy, lensing surveys have to
make very accurate determinations of galaxy shapes by correcting for
the smearing effect of the Point Spread Function (PSF). 
The PSF determination is obtained generally from stars which have a
bluer spectrum than faint distant galaxies. In addition,
in galaxies, the spectrum can vary with the position in the galaxy image.
Color gradients within galaxies can thus bias the shape measurements. 
Voigt et al.~(2011) estimated this bias out to $z\sim1$ by modelling galaxies using two
co-centered, co-elliptical Sersic profiles, each with a different
spectrum. They showed that the shear bias from a single galaxy can be 
substantial depending on the properties of the galaxy and the width of
the filter used. Using a small galaxy catalog from
Simard et al.~(2002) which was based on WFPC2 observations of the `Groth
Strip' (Groth et al.~1994, Rhodes, Refregier \& Groth 2000) and which goes up to
$z\sim1$ (median redshift of 0.65), they
found a mean shear bias that is lower than the statistical errors for
future cosmic shear surveys. However, the true mean shear bias may
exceed the statistical errors depending on how accurately this
catalog represents the observed distribution of galaxies in cosmic
shear surveys. In this paper, we use a large galaxy sample (3248
objects down to $i_{AB}=25.5$ and $z\sim3$) in the GOODS-South field (Giavalisco et al.~2004) 
to characterize the color gradients in galaxies and their dependence
on galaxy properties. The determination of these gradients for a large
statistical sample will enable corrections to be made to this systematic
in shear measurements. 

The measurement of these gradients also has implications
for the evolution of galaxies. By stacking galaxies in volume-limited
subsamples, we also characterize the evolution of the radial color
variation with redshift, from the local Universe out to $z\sim3$.
The relative difference in color between the interior of the galaxy
and the outskirts, and its evolution with galaxy properties and in
particular with redshift, indicates whether star formation in these
galaxies occurs preferentially in the nucleus or the outskirts of the
galaxies and if this location changes on average with redshift and
other galaxy properties. Several studies of spatially resolved galaxy
properties, including star formation, at $z\sim0.1$ have been performed by Welikala et
al.~(2008; 2009) and Park et al.~(2007) in the Sloan Digital Sky
Survey (SDSS; Abazajian et al.~2009). These studies have explored the
relation between spatially resolved color (and hence stellar
population) trends and the galaxy environment. Tortora et al.~(2010; 2011) explored both color gradients and
stellar M/L gradients in galaxies in SDSS and
showed that M/L gradients are strongly correlated with color
gradients. Suh et al.~(2010) studied early-type galaxies drawn from
the SDSS DR6 and found a strong correlation between the existence of
steep colour gradients and ongoing residual star formation. They also
showed that elliptical galaxies with bluer cores had global bluer colors than average. Lee et al.~(2008) found
that steeper colour gradients appear 
within star forming galaxies, in both late and early-types.
Color gradients of low-redshift spiral galaxies appear to be dominated by the fact
that, on average, their bulges are redder than their disks. 
However, this may be an oversimplification for many late-type galaxies (e.g., Bakos et al.~2008).
Gonzalez-Perez et al.~(2011) studied the relation between 
color gradients in galaxies at $0.01<z<0.17$ in SDSS DR7 and their
galactic properties. They found that, on average, galaxies in this
redshift range have redder cores than their outskirts and found a
steeper gradient in late-type galaxies than in early-types. They also
demonstrated a connection between steep color gradients and a higher
proportion of young stars within a galaxy, and found that nuclear
activity is a marginal driver for creating steep color gradients in
massive galaxies. However, the relation between color gradients and
galaxy properties at higher redshifts have not been fully explored to date with large
galaxy samples.

Some studies at higher redshifts have been performed using relatively small samples
of galaxies. For example, Ferreras et al.~(2005) studied the evolution
of 249 field early-type galaxies in GOODS-South, with a median
redshift of 0.71 and measured their color gradients. They found that red and blue early-type galaxies in
their sample have distinct behavior with respect to their color
gradients. In particular, they found that most blue early-type
galaxies feature blue cores whereas most red early-types have
passively evolving stellar populations with red cores, i.e., similar
to local early-type galaxies. Furthermore, they found that, out to
$z\sim1$,  color gradients and their scatter do not evolve with
redshift and are compatible with the observations at $z\sim0$,
assuming a radial dependence of the metallicity within each
galaxy. They also ruled out significant gradients in the stellar age
out to $z\sim1$. Brok et al.~(2011) used deep, HST imaging to determine
color profiles of early-type galaxies in the Coma cluster. They found
negative color gradients for these galaxies and in addition, found that color profiles are typically linear as a function of $log(R)$,
sometimes with a nuclear region of distinct, often bluer color. They
found that color gradients of dwarf galaxies form a continuous
sequence with those of elliptical galaxies, becoming shallower toward
fainter magnitudes. Using the colors as metallicity tracers, they
suggested that dwarfs as well as giant early-type galaxies in the Coma
cluster are less metal rich in their outer parts. They did not find
evidence for the environment of galaxies impacting the gradients, although most
of the galaxies in their sample were found in the central regions of
the cluster. They also found that S0 galaxies in a subsample have less steep gradients than elliptical galaxies.

%[HERE IT IS ESSENTIAL WE EXPLAIN THAT OUR AIM HERE. WE ARE MEASURING COLOR GRADIENTS
%AS WOULD BE OBSERVED IN A LENSING SURVEY, SO WE ARE NOT PARTICULARLY
%CONCERNED WITH TRACING A PARTICULAR POPULATION OF GALAXIES WHICH WOULD
%REQUIRE CARFUL SELECTION. WE HAVE TO BE REALLY CAREFUL IN WORDING THIS
%I THINK. NEED YOUR INPUT HERE.]

In this work, we are interested in the impact of color gradients 
out to higher redshifts ($z\sim3$) as would be observed by future deep
and wide cosmological weak imaging lensing surveys such as the Dark
Energy Survey (DES\footnote{\url{http://www.darkenergysurvey.org}}; Wester et
al.~2005), the Kilo-Degree Survey (KIDS\footnote{\url{http://www.astro-wise.org/projects/KIDS/}}), the
Hyper Suprime-Cam (HSC) Project\footnote{\url{http://www.naoj.org/Projects/HSC/HSCProject.html}}, the Large Synoptic Survey Telescope (LSST\footnote{\url{http://www.lsst.org}}; LSST Collaboration 2009),
\textit{Euclid}\footnote{\url{http://www.euclid-ec.org}} (Laureijs et
al.~2011) and the Wide-Field InfraRed Survey Telescope (\textit{WFIRST}; Green
et al.~2011). This relies on a
large photometric galaxy sample out to high redshifts. The lensing shear
measurements will be impacted by galaxies of varying types,
luminosities, sizes and redshifts, and the aim of this paper is to
explore for the first time these dependences that will affect the
shape measurements of galaxies. In particular, lensing measurements 
and the impact of color gradients on these measurements have been
studied in the context of the SNAP (SNAP Collaboration 2004) /JDEM project (Seiffert, private
communication) and the \textit{Euclid} project (e.g Voigt et
al.~2011). For this reason, the \textit{Euclid} Visible Channel (VIS) is also equipped
with a narrow band filter that will enable the determination of the
correction of the color gradient effect as a function of galaxy properties
and redshift out to $z\sim3$. 
%In this work, we aim to assess the possible impact of color gradients on shear
%measurements out to $z\sim3$. This is particularly relevant for future
%mission such as \textit{Euclid} and \textit{WFIRST} which would make such measurements on a representative sample of galaxies out to $z\sim3$.

However, by exploring these trends, we also show trends in the
color gradients that are directly related to star formation and
possibly other astrophysical processes in galaxies. By stacking
objects in volume-limited samples within our large sample, we also aim 
to investigate the evolution of the color gradient from the local
Universe out to $z\sim3$. This is likely to have an interpretation in terms of how
star formation and feedback in galaxies evolve across cosmic time.
We use the AB magnitude system throughout this paper.

\section{The photometric sample}
\label{sec:specdata}

We use a magnitude-limited ($i<26$) sample of galaxies in the GOODS-South
survey (Giavalisco et al.~2004) from version 2.0 of the publicly available ACS
source catalog\footnote{\url{http://archive.stsci.edu/pub/hlsp/goods/catalog_r2}}
as well as a photometric redshift galaxy catalog for CDFS (Cardamone
et al.~2010). The ACS source catalog is produced using the SExtractor
package (Bertin \& Arnouts 1996). This also
determine the position of the central pixel of each galaxy. 
We then produce a $5\arcsec\times5\arcsec$ postage stamp cutout image of
each galaxy from the reduced,
calibrated, stacked and mosaiced ACS images\footnote{\url{http://archive.stsci.edu/pub/hlsp/goods/v2/}} in the $v606$ and $i775$ passbands. We also apply a selection on galaxy size, as determined by tests described in
Section 3.2. This leaves us with a photometric sample of 3248 galaxies with $v$ and $i$
band images.

In order to compare any systematic effect in the dependence of the
color gradient on redshift due to the use of photometric redshifts as opposed to spectroscopic redshifts, we also
select a subsample of the galaxies from our photometric catalog which fall in the GOODS-VIMOS
spectroscopic campaign (Popesso et al.~2009; Balestra et
al.~2010). This consists of two surveys which 
target galaxies in different redshift ranges.  
The VIMOS Low Resolution Blue (LR-Blue) is aimed at observing
galaxies mainly at $1.8<z<3.5$ while the Medium Resolution (MR) orange
grism is aimed mostly at galaxies at $z<1$ and Lyman Break Galaxies
(LBGs) at $z>3.5$. These leave 531 galaxies with both spectroscopic
and photometric redshifts for the comparison. 

Figure 1 illustrates the selection of both samples in
luminosity and redshift as well as in apparent magnitude and half light radius. The photometric
sample is complete down to $z\sim2.5$ and $M_I < -20.0$, while the
spectroscopic sample is complete down to $z\sim2$ and $M_I<-21.0$. 
The magnitude limit of the photometric sample is 25.5 while it is 24.5
for the spectroscopic sample. 

Figure 2 illustrates the distribution of each of the galaxy properties
being studied in the sample. In terms of photometric
redshift, there are 638 galaxies at $z<0.5$, 1195 galaxies
at $0.5<z<1.0$, 768 glaxies at $1.0<z<1.5$, 320 galaxies at
$1.5<z<2.0$. In particular, there are 277
galaxies in the redshift interval $2.0<z<3.0$, allowing a statistically
significant evolution of the color gradient between $z=2$ and $z=3$. 
The photometric redshift distributions of the early and late-type
galaxy samples are also shown in Figure 2.
The distribution of spectroscopic redshifts is also included. There
are 109 objects in the spectroscopic sample in the redshift interval
$2<z<3$, with a median redshift of 0.84. 

1171 galaxies have $i_{AB}>24.0$. 594 galaxies are in the absolute
magnitude range $-21.0<M_{I} \le -20.0$ (henceforth, 'faint') and 
679 galaxies in the $-22.0<M_{I} \le -21.0$ (henceforth, 'bright').
Futher, we investigate morphological dependences of our results by
splitting our sample into early and late-type galaxies using a publicly available morphology catalog for
GOOD-South\footnote{\url{www.ugastro.berkeley.edu/~rgriffit/gems_v_z_public_catalog_5.0.fits}}
(Griffiths et al. 2012, in prep). The sample is dominated by late-type galaxies
($n_{Sersic} < 1.5$) accounting for 1938 objects, 6 times the number
of early-type galaxies ($n_{Sersic}>2.5$). 1915 galaxies in the sample
have  $v-i<1$. The mean half light radius ($R_{50}$) of the sample is
0.45$\arcsec$. The mean half light diameter is thus 7 times the
angular size of the resolution element (PSF FWHM in the i band is $\sim0.11\arcsec$).

\section{Methods}

\subsection{Measuring the color gradients in galaxies}

ACS science images are in counts/sec, while the weight maps
are inverse variance maps (i.e., $1/\sigma_{f}^2$).
Flux calibration of the science and weight postage stamp
ACS images of each source is peformed using the ACS  zeropoints
listed\footnote{\url{http://archive.stsci.edu/pub/hlsp/goods/v2/h_goods_v2.0_rdm.html}}. These
contain the variance from the sky noise, read-out noise and dark
currents. 

Futher, because the pixels are correlated, the actual noise is higher than the
theoretical noise in the weight files (Casertano et al.~2000). We account for this 
by computing the noise in empty patches (away from the object and
whose size is larger than the correlation length scale) in the science
image and we calculate the ratio of this measured noise value to the
theoretical noise value. The mean ratio is 0.6. The noise in the weight images is then rescaled
by this ratio to give a more accurate estimate of the noise in the
pixels due the background and instrument. The Poisson uncertainty from
pixels in the source itself is then added.

The $v$ ad $i$ images are PSF-matched using
the \textit{ip\_diffim}\footnote{\url{http://dev.lsstcorp.org/trac/}}
image mapping software currently in the pipeline of LSST. This a version of the 
Higher Order Tranform of Psf and Template Subtraction code
(\textit{Hotpants}\footnote{\url{http://www.astro.washington.edu/users/becker/hotpants.html}}). This
code implements the algorithm of Alard (1999) and Alard \& Lupton (1998) for
image subtraction by finding the kernel that
is needed to map a input image to a template image. The kernel is
decomposed into a set of basis functions, usually
 Gaussians of varying FWHM. The process is thus a linear least-squares
 problem and can be solved via matrix inversion. This process matches
 the PSFs of the two input images given an appropriate kernel. Since the PSF
varies spatially in all the images, the code models the kernel as a
spatially varying function as well. A bounding box is chosen for each
galaxy, centred on the central pixel of each galaxy (as found above)
in order to avoid masked or bad pixels from affecting the procedure. A similar process is used to PSF-match the
variance images.

The color gradient across the galaxy is measured as:

\begin{equation}\label{first}
      \delta(v-i)= (v-i)_{R_{50}\leq r<2R_{50}} - (v-i)_{r<R_{50}} \\
\end{equation}

where $(v-i)_{r<R_{50}}$ is the $v-i$ color inside $R_{50}$
and $(v-i)_{R_{50}\leq r<2R_{50}}$ is the $v-i$ color between  $r\geq R_{50}$ and $r<2R_{50}$ 
The upper limit of $2R_{50}$ is validated by determining the radial
dependence of the signal-to-noise ratio (SNR)  in volume-limited subsamples
of galaxies out to $z\sim3$. In galaxies at $z\sim2.5$,  
the SNR declines significantly beyond $2.5R_{50}$.
In addition, when these galaxies at $z\sim2-3$ are stacked, the SNR
in the annulus $1.5R_{50}<r<2R_{50}$ is 9.0 and falls below 2.0 
for $r>2.5R_{50}$. See Section 4.3 for more details.

Photometric errors in each pixel are propogated from the variance
images in order to provide an
uncertainty on the measurement of the color gradient for each galaxy.

\subsection{Effect of galaxy size on the measured color gradient}

We perform simulations to determine how the input galaxy size in these
objects impact the accuracy with which a color gradient can be
recovered for these galaxies. We simulate mock disk
galaxies with an exponential profile at a given redshift $z$ and with
a range of values of $R_{50}$.  Luminosities are assigned to each particle in the
disk using SEDs generated from the Bruzual \& Charlot 2003
(BC03) stellar population synthesis models. These luminosities are
assigned to the stellar particles such that an input color gradient
exists in the galaxy between $r<R_{50}$ and $R_{50}<r<2\times R_{50}$. Mock images in the
ACS $v$ and $i$ filters are made by
redshifting the SEDs to the redshift of the galaxy and convolving the SEDs with the
ACS filters. The images are then pixelated to the ACS pixel scale
(0.03''/pixel in the drizzled images) and smoothed with a Gaussian
kernel of FWHM 3.7 $\times$ 3.7 pixels in order to simulate the resolution
of the actual ACS images (PSF FWHM in i band $\sim0.11\arcsec$). 

The results of this simulation are shown in Figure 3 where we plot the fractional error in
the recovered color gradient $v-i$ as a function of the input $R_{50}$
of the galaxy for a range of color gradients which are assigned initially to the
galaxy. As expected, the
fractional error increases as the galaxy size decreases. However, for
$R_{50} < 0.3\arcsec$, the fractional error exceeds 0.2 and increases
sharply beyond this value. We thus use $R_{50}>0.3\arcsec$ (half light
diameter $\sim 5.5 \times$
FWHM $PSF_{i}$) as a benchmark to select galaxy sizes in our sample. 
Note that this size limit is more conservative than the selection of
$d >1.6 \times FWHM$ ($d$ being the typical galaxy size) in the COSMOS
survey (Scoville et al.~2007; Koekemoer et al.~2007) for galaxy 
shape measurements for weak lensing (Leauthaud et al.~2007).

%We aso test the impact of the uncertainty in the magnitude of the
%object in both filters on the recovered color gradient. It is clear
%that for $\delta\,mag_{i,v} > 0.2$ magnitudes, the fractional error in the color gradient
%deteriorates significantly. 

\section{RESULTS}
\label{sec:results}

\subsection{Observed color gradients and their dependence
  on galaxy properties}

Figure 4 illustrates the dependence of the observed color gradient on 
galaxy properties. We observe the color gradient increasing with 
both the $v$ and $i$ band apparent magnitudes out to $\sim25.5$ magnitudes. 
This implies that fainter galaxies are progressively redder in their
outskirts than in their centres, while brighter galaxies have a
negative color gradient, implying that their outskirts are
bluer than their centres. The color gradient also increases rapidly for galaxies fainter
than $i=22.5$ magnitudes. A similarly large, though less steep rise in
the color gradient is observed in the
$v$ band magnitude. We observe a decreasing color gradient with increasing 
$R_{50}$ i.e.~small galaxies have bluer cores compared to their
outskirts, but as the galaxy size increases, the core becomes
progressively redder compared to the outskirts, resembling local
early-type galaxies more closely. 

The trend with redshift is also very remarkable. There is little evolution in
the color gradient or in its scatter out to $z\sim1$. 
However, beyond $z\sim1$, there is a sharp increase in the color gradient out to $z\sim3$,
implying that galaxies become bluer in their core relative to their
outskirts. This can be interpreted as follows. For $z<1$, we are
measuring rest-frame optical colors which probe older stellar populations. For
$z>1$, we are measuring rest-frame UV colors which are more
sensitive to the bluer and younger stellar populations.  In particular, we
observe bluer colors in the centres of galaxies relative to
the outskirts above $z\sim1$. There is an indication also that the
 increasing trend flattens out from $z\sim2$ to $z\sim3$. The redshift evolution of the
median color gradient for $0.5<z\leq3.0$ can be described by the parametric function
$\delta_{(v-i)}(z) = -0.033z^3 + 0.119z^2 + 0.045z - 0.093$ (a bin
size of $\triangle{z}=0.5$ is used, with the data points in Figure 4
corresponding to the centres of the bins). There is also a strong dependence of the gradient on $(v-i)$ color. The bluest galaxies
 have a positive color gradient $\sim0.25$ magnitudes i.e.\ their
 cores are bluer relative to their outskirts, while redder galaxies
 have redder cores relative to their outskirts, consistent with local 
S0/Sa galaxies. The trend with absolute magnitude is weak although
there is an indication that the color gradient increases and its scatter decreases
slightly towards the lowest luminosities. Finally, we examine the
trend of the gradient with morphological
type. There is not a strong dependence but a suggestion of a slight
decrease in the color gradient with increasing Sersic index. For $n<1.5$ (late-type
galaxies),  there is an indication that these
galaxies have bluer centres compared to early-type galaxies ($n>2.5$), but the trend is
quite weak for the whole sample. We examine the dependence of the
color gradient on galaxy morphology in more detail below.

\subsection{Dependence on morpological type}

In order to investigate any dependence of galaxy morphology on
the results, we perform a similar analysis on late-type galaxies
($n_{Sersic} < 1.5$) and early-type ($n_{Sersic} > 2.5$) in the sample. The results are shown in Figure 5. 
The trends of $\delta_{(v-i)}$ with galaxy properties for late-type galaxies are
very similar to the trends in the full galaxy sample since the latter
is dominated by late-type galaxies. The trends are also qualitatively similar between early and late-type
objects but there are some notable differences. The color gradient for late-type
galaxies increases monotonically from the brightest to the faintest
magnitudes ($i_{AB}\sim25.5$). The trend with $i_{AB}$ apparent magnitude for
early-type galaxies shows some differences compared to late-type
galaxies. The median color gradient for
early-type galaxies shows very little variation with magnitude up to $i_{AB}=22.5$ but the relation also has a
larger scatter than for late-types. Beyond $i_{AB}=22.5$, the median $\delta_{(v-i)}$ increases sharply up to 0.15 magnitudes at $i_{AB}=23.5$, indicating
that the cores of early-type galaxies become bluer compared to their
outskirts. Beyond $i_{AB}=23.5$, the median $\delta_{(v-i)}$ shows little variation
with $i_{AB}$ magnitude. The variation of the color gradient in early-type galaxies 
with the $v_{AB}$ magnitude is similar to that in late-types, with
both galaxy types showing little variation of $\delta_{(v-i)}$ with $v_{AB}$ up
to $v_{AB}=23.5$. However, the scatter in the relation at the faintest magnitudes ($i_{AB}>23.5$)
is much larger for early-type galaxies. 
 
 There is some difference in
the evolution of the color gradient with color. Bluer early-type galaxies ($v-i<1.0$)
have a color gradient which shows little variation with color but for $v-i>1.0$, $\delta_{(v-i)}$ decreases towards redder early-type objects. The bluest
early-type galaxies have a median $\delta_{(v-i)}$ that is closer to zero indicating that
the bulge and disk components are not as well separated as for
late types. By comparison, the bluest late-type objects have cores
which are bluer by as much as 0.25 magnitudes compared to their 
outskirts, and $\delta_{(v-i)}$ then decreases monotonically with $v-i$ color.
Both the reddest early-type and late-type galaxies have cores which are redder
by as much as 0.15 magnitudes compared to their outskirts. 

The color gradient relation with $R_{50}$  has a larger scatter for early-type
galaxies than for late-types. There is an indication of a reduced
color gradient for larger early-type galaxies and this trend is also observed for
late type galaxies although the scatter in the relation is larger for
early-type objects. These
objects with large $R_{50}$ are probably similar to local early-type
galaxies that have red cores.
Late-type galaxies also show little or no evolution in the color gradient with luminosity (as in the case of
the full sample) but early-type galaxies do show an indication that
the gradient increases, from the
intrinsically brightest objects at $M_{I}=-23.0$ which show bluer outskirts, to the lowest luminosities
where galaxies have slightly redder outskirts compared to their centres. 

The trends with redshift are broadly similar for early and late-type
galaxies, although the scatter in the trend for the early-type
galaxies is larger. Early-type objects show little or no evolution in
their color gradient below $z\sim1$, consistent with the results of Ferreras et
al.~(2005). However, for $z>1$, their color gradient becomes increasingly positive
before flattening off around $z\sim1.5$. This implies that early-type
galaxies become bluer in their centres at high redshifts, in marked contrast to
their local counterparts. Finally, in the redshift interval
$z=2.5-3.0$, $\delta_{(v-i)}$ is +0.17 for early-type galaxies and +0.24 for
late-type galaxies. This indicates that at high redshifts, the cores
of late-type objects not only become progressively bluer relative to
their outskirts but that the difference between the color of the
outskirts and the color of the core becomes stronger in late-type galaxies than it
does for early-type galaxies.

\subsection{The stacked radial color variation in galaxies}

We investigate further the observed evolution of the color gradient with redshift by
measuring the radial varation of color in two luminosity
ranges $-21.0<M_{I} \le -20.0$ and $-22.0<M_{I}
  \le -21.0$. We measure the $v-i$ color in successive radial annuli for
each galaxy in each absolute magnitude interval. 
%Since galaxies with the smallest $R_{50}$ are mor at higher
%redshifts, 
We select galaxies in the range $0.30<R_{50}<0.40$ in order
to verify the measurement accuracy, since these galaxies will be 
reprensentative of objects at the highest redshifts. Galaxies in each of these absolute magnitude intervals
  have the radius of their annuli scaled by $R_{50}$ before the 
  galaxies are stacked. We have performed tests to ensure that our outermost annuli are not
biased by the color of the background. For $z\sim2.0-3.0$, the SNR for the stacked galaxies in the
outermost annulus ($r=1.5-2R_{50}$) is 9.0, and is always above 2.0 for any radius below
2$R_{50}$. Beyond $2R_{50}$, the SNR decreases sharply below 2.0. 

The results are shown in Figure 6. In both luminosity intervals, we
observe the color gradient becoming progressively more positive for $z>1$ i.e.,
showing increasingly bluer cores relative to the outskirts. 
At $z=0.5-1.0$, the cores of galaxies ($r<0.5R_{50}$) are slightly redder on average
 compared to their outskirts by 0.05-0.1 magnitudes in both luminosity
 intervals. However, at $z=2.5-3.0$ and $-21.0<M_{I} \le -20.0$, the core is bluer by 0.6
 magnitudes (at a 16$\sigma$ significance level) compared to the
 outskirts ($r=1.5-2R_{50}$). In the intrinsically brighter
 galaxies ($-22.0<M_{I}\le -21.0$) in the same redshift interval, the core is bluer by 0.42
 magnitudes compared to the outskirts. We remark that the radial
 trends are qualitatively similar in both luminosity intervals although the exact magnitude of the
 color difference between the core and the outskirts can vary. 

%The color gradient at $z<1.25$ is therefore zero or slightly
%negative but becomes increasingly positive with redshift for $z>1.25$. 

%There is little change in the color gradient between the two luminosity
%bins, consistent with the very weak trend of $\delta_{(v-i)}$ with $M_{I}$ observed in
%Figure 4. 
%At $z\sim2$, the brighter galaxies are bluer
%at all radii by as much as 0.25 magnitudes. 
Another very important point to emphasise is that for both luminosity ranges, the color in
the outskirts of the galaxy changes much less than the internal color.
The color at $r\sim1.5-2.0R_{50}$  changes by
approximately 0.2 magnitudes between $z\sim0$ and $z\sim3.0$, for
galaxies in $-22.0<M_{I}\le -21.0$ (and 0.05 magnitudes in $-21.0<M_{I}\le -20.0$). 
In comparison, in the interior ($r<0.5R_{50}$), the color changes by as much as 0.77
magnitudes (0.57 magnitudes at $-21.0<M_{I}\le -20.0$). 
This indicates that as the star formation rate in galaxies increases
towards $z\sim3$ (galaxies becoming bluer), it is the central regions
of galaxies where this increase is most marked, possibly due to increased gas
densities in the central regions.

\subsection{Biases in the color gradient arising from spectroscopic redshifts}

In Figure 7, we investigate any bias in the evolution of the color gradient when
using spectroscopic
redshifts as opposed to photometric ones. 
The trend of the color gradient with the spectroscopic redshift is similar
to the trend with the photometric redshift.  However, we find that, up
to $z\sim1$, the median color gradient for the
spectroscopic sample is lower by 0.04 magnitudes compared to the
photometric sample and it is lower by 0.07 magnitudes at $z\sim2$. This can be explained by the
different color distributions of the two samples, with the spectroscopic
sample (which is color-selected) being redder by 0.2 magnitudes than the photometric one at
$z\sim1$. This bias is also consistent with the strong color dependence we
found for the color gradient (Figure 4), with redder galaxies having
a less positive color gradient.

%However, beyond $z\sim1$, 
%it is slightly offset compared to the photometric redshfit sample 
%towards a negative color gradient. This implies that the spectroscopic
%redshifts are sampling an underlying poulation of galaxies with 
%bluer outskirts (and redder cores) than does the photometric sample.
%We investigate this further by examining the dependence of the $v-i$
%color on the photo-$z$ or $z_{spec}$. This is illustrated in the
%bottom panel of Figure 4. 

%TALK ABOUT PSF CORRECTION

\section{Discussion}
\label{sec:conclusion}

\subsection{Interpretation in terms of galaxy evolution}

The observed evolution with redshift is consistent with the picture of galaxies being
increasingly star-forming at higher redshifts. Importantly, our
results suggest that the star
formation becomes increasingly dominant in the cores of galaxies
relative to their outskirts, at higher redshifts. This is
consistent with the star formation being concentrated in the nuclei of galaxies, possibly
driven by higher gas densities in the centre. 
%Green et al.~2010 found a trend
%between the SFR and the $H{\alpha}$ linewidths of galaxies at $z\sim0.1$. 
There are several studies that support the picture of very high star
formation rates in $z\sim1-3$ galaxies being driven by high gas
densities and self-regulated star formation i.e., the star formation
determines the average thermal and turbulent pressure in the
inter-stellar medium (ISM), which in turn affects the star
formation rate. Lehnert et al.~(2009) have proposed that the large $H\alpha$
line widths observed in $z\sim1-3$ intensely star-forming galaxies are
driven by self-regulated star formation through the mechanical energy
liberated by massive stars. 
%They proposed
%that this mechanical energy is sufficient to keep the disk critically
%unstable against fragmentation and collapse. 
They also found that the intensity of the star formation in these
distant galaxies is as high as that observed in local starbursts
(e.g., M82), but the star formation occurs on a much larger
physical scale, and it is maintained by large gas fractions and high
mass-surface densities. These $H\alpha$ line widths do not appear to be driven
by either cosmological accretion (Le Tiran et al.~2011a) or
gravitational instabilities. 

Further, Le Tiran et al.~(2011b) also stacked rest-frame optical emission
lines such as [SiII]$\lambda\lambda$6716,6731 and [OI]$\lambda$6300,
of about 50 color-preselected galaxies at $z=1.2-2.6$ which also had high $H\alpha$
surface brightnesses, in order to investigate how gas properties scale
with the star formation intensity. They found that their high redshift
sample showed trends similar to those observed in local galaxies where gas pressures
scale with the star-formation intensity. In particular, they found higher gas densities (and hence
pressures) in intensely star-forming regions compared to fainter
diffuse gas, with values comparable to starburst regions and the
diffuse ISM in nearby galaxies. In their stacked sample, they found that in
addition to single narrow components of $H\alpha$ and
[NII]$\lambda\lambda$6548,6583, broad lines in $H\alpha$ and
[NII] were also observed. These broad lines are only significantly detected in the stacks with the
highest star-formation intensities. This broad emission supports the hypothesis that
outflows, rather than active galactic nuclei, are closely linked to the high
star-forming intensities. 

Thus, if very high gas densities are indeed present in the centres of $z>1.5$
galaxies, this would support the hypothesis of very strong star
formation which is driven by feedback and self-regulation and is concentrated in the centres of galaxies as opposed to their
outskirts. That would be consistent with the color gradients we
observe at these high redshifts. 

%In comparison, below $z\sim1$, star formation in galaxies appears to
%be more uniformly distributed through the disks.

\subsection{Implications for shear measurements}

The observed trends in the color gradient discussed above are
important in the context of weak lensing galaxy shape
measurements. Indeed, the faint distant galaxies targeted by weak
lensing surveys are convolved by the instrumental PSF which is a
function of wavelength (generally being broader at longer
wavelengths). Hence, galaxies of a given intrinsic shape (and color) but with
different color gradients will have different observed
shapes. Uncovering the original galaxy shape thus requires, in
principle, a knowledge of the PSF variation with wavelength (as
studied by Cypriano et al.~2010) and also of the galaxy's intrinsic color gradient.

In the case of ground-based observations such as those conducted with
CFHT/Megacam, the new wide-field imager VST/OmegaCam, and the
upcoming DES and HSC projects, the limited spatial resolution may not allow an
identification of the color gradients of galaxies. However, the PSF
of ground-based observations is well characterized in different bands,
and an accurate measurement of integrated galaxy colors can be made. The variation of the color
gradient with galaxy color and redshift that we have identified in
this paper, should allow us to build a model for correcting the PSF smearing by including the galaxy color and photometric redshift in the shape measurement scheme.

In the case of space-based observations, color gradients can be easily measured 
when multi-band information is available, as demonstrated here, so
modelling the correction of the PSF smearing should not be a difficult problem for resolved multi-band observations. 
%However, for future experiments where only one single (very broad)
%filter is to be considered, an accurate PSF correction might 
%be biased by the intrinsic galaxy color gradients.
%Therefore a dedicated evaluation is required in order to investigate how
%significant such a hardware limitation is. 
These observations can be used to 
quantify the size of the bias as a function of galaxy properties,
which in turn can be used to correct the bias. In a forthcoming paper, we will investigate 
these different issues for both the ground and space-based observations. 
We will propose a model that could potentially correct the shape 
measurement bias, taking into account the variation of the color gradient with galaxy properties as presented in this work.

\section{Acknowledgements}
NW thanks L. Tasca for provided the cutout images for this
analysis, O. Ilbert for useful discussions and A. Becker and R. Owen for
help and advice on using code from the LSST Trac. We thank Y. Mellier
and H. Hoekstra for their helpful suggestions. NW acknowledges support from
the Centre National d'\'Etudes Spatiales (CNES) and the Centre National de
la Recherche Scientifique (CNRS). JPK acknowledges support from CNRS
and CNES.

\bibliographystyle{plain}

%%%%%%%%%%%% Figures %%%%%%%%%%%%%%%%%%%

\begin{figure}
\centerline{\rotatebox{0}{
\includegraphics[height=0.45\textwidth]{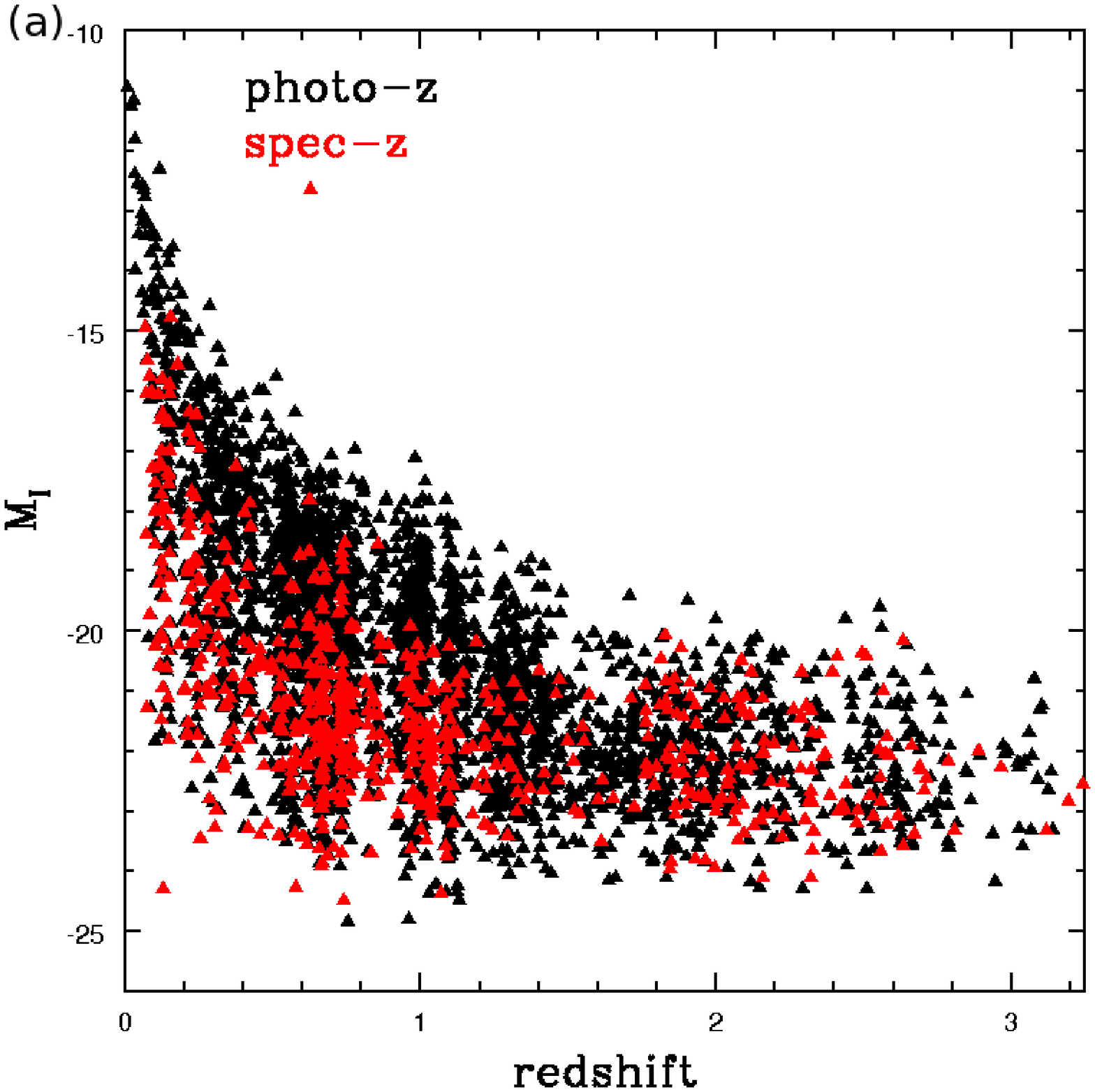}
%\hspace{0.1cm}
%\includegraphics[height=0.45\textwidth]{z-IAB.eps}
}
}
\vspace{0.1cm}\centerline{\rotatebox{0}{
\includegraphics[height=0.45\textwidth]{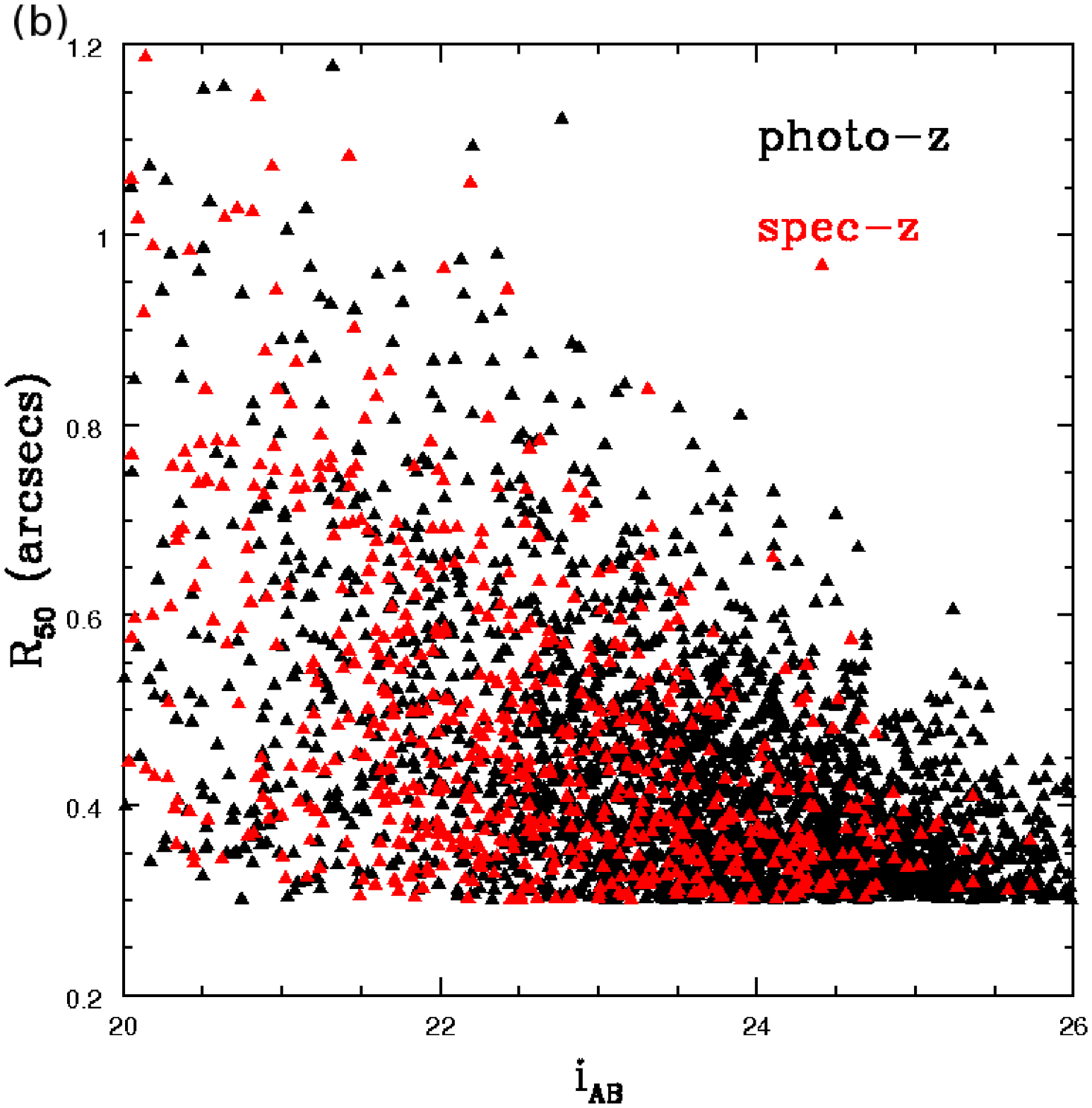}
}
}
\tiny{
\caption{Selection of the photometric (in black) and spectroscopic
  samples (red) in GOODS-South that were used in this study. (a)
  $M_I$ as a function of redshift and (b) $R_{50}$ versus
  magnitude.
}
\label{fig:selection}}
\end{figure}

\begin{figure}
\centerline{\rotatebox{0}{
\includegraphics[width=0.35\textwidth]{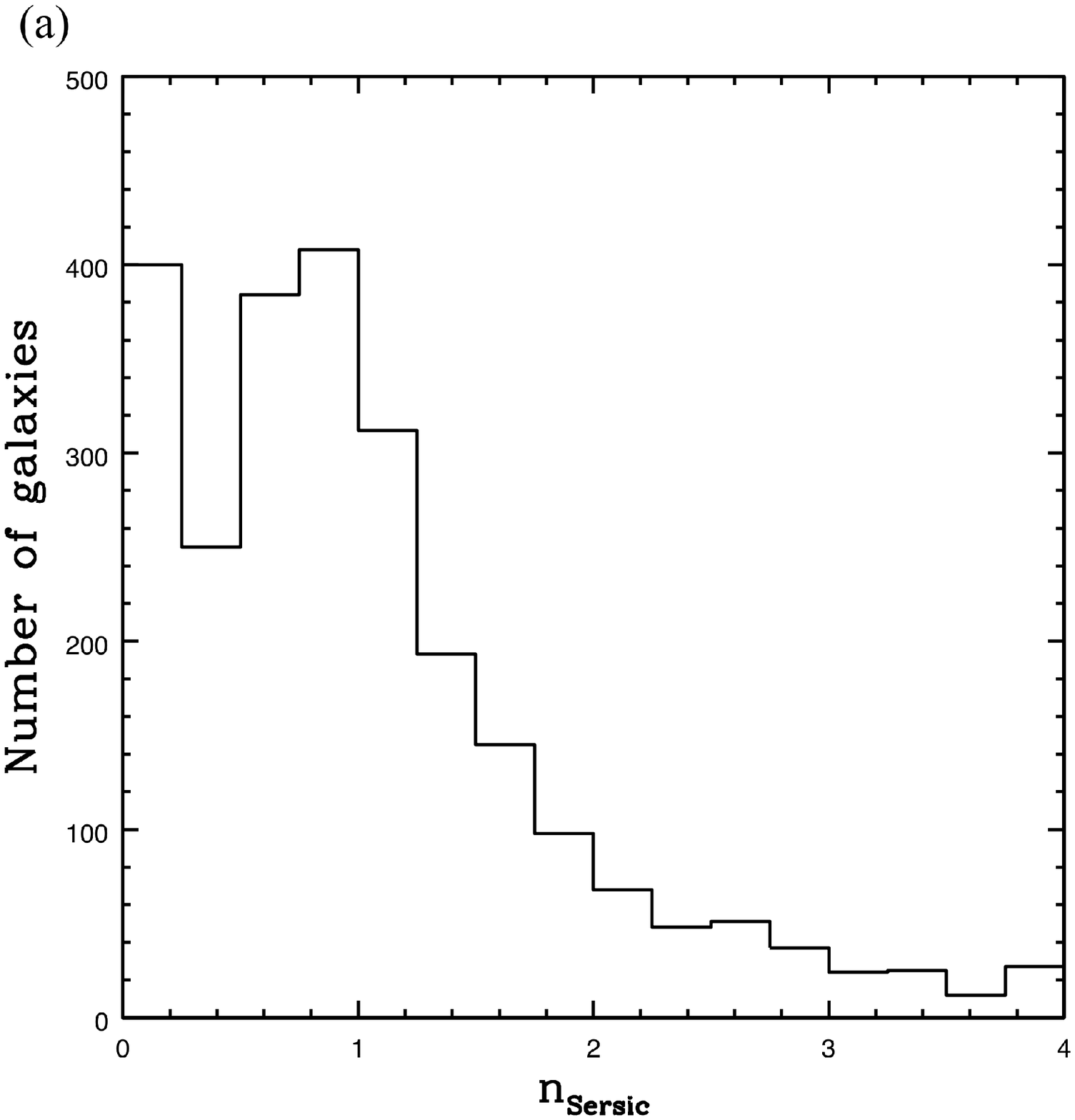}
\hspace{0.1cm}
\includegraphics[width=0.35\textwidth]{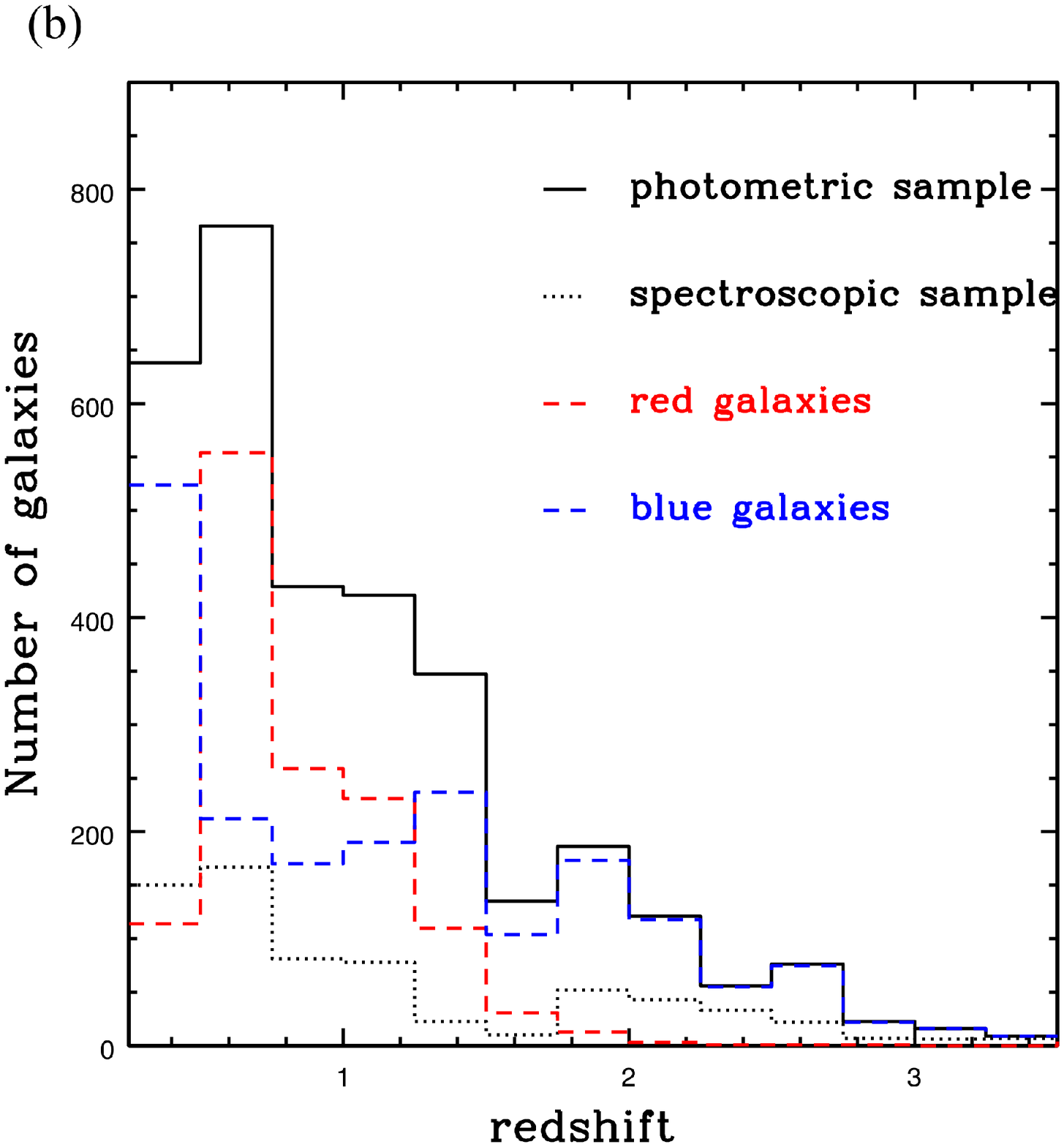}
\hspace{0.1cm}
\includegraphics[width=0.35\textwidth]{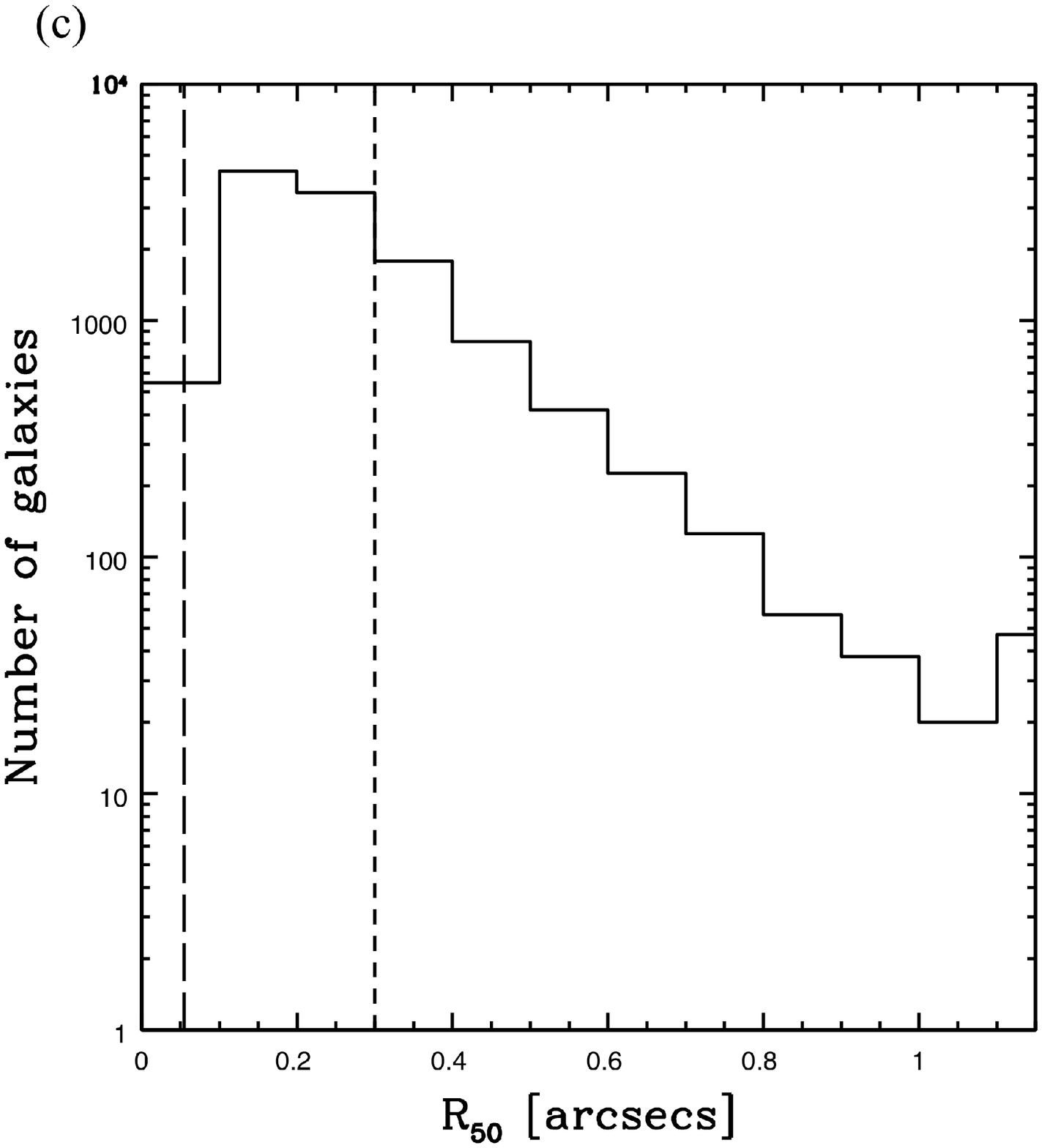}, 
}
}
\vspace{0.1cm}\centerline{\rotatebox{0}{
\includegraphics[width=0.35\textwidth]{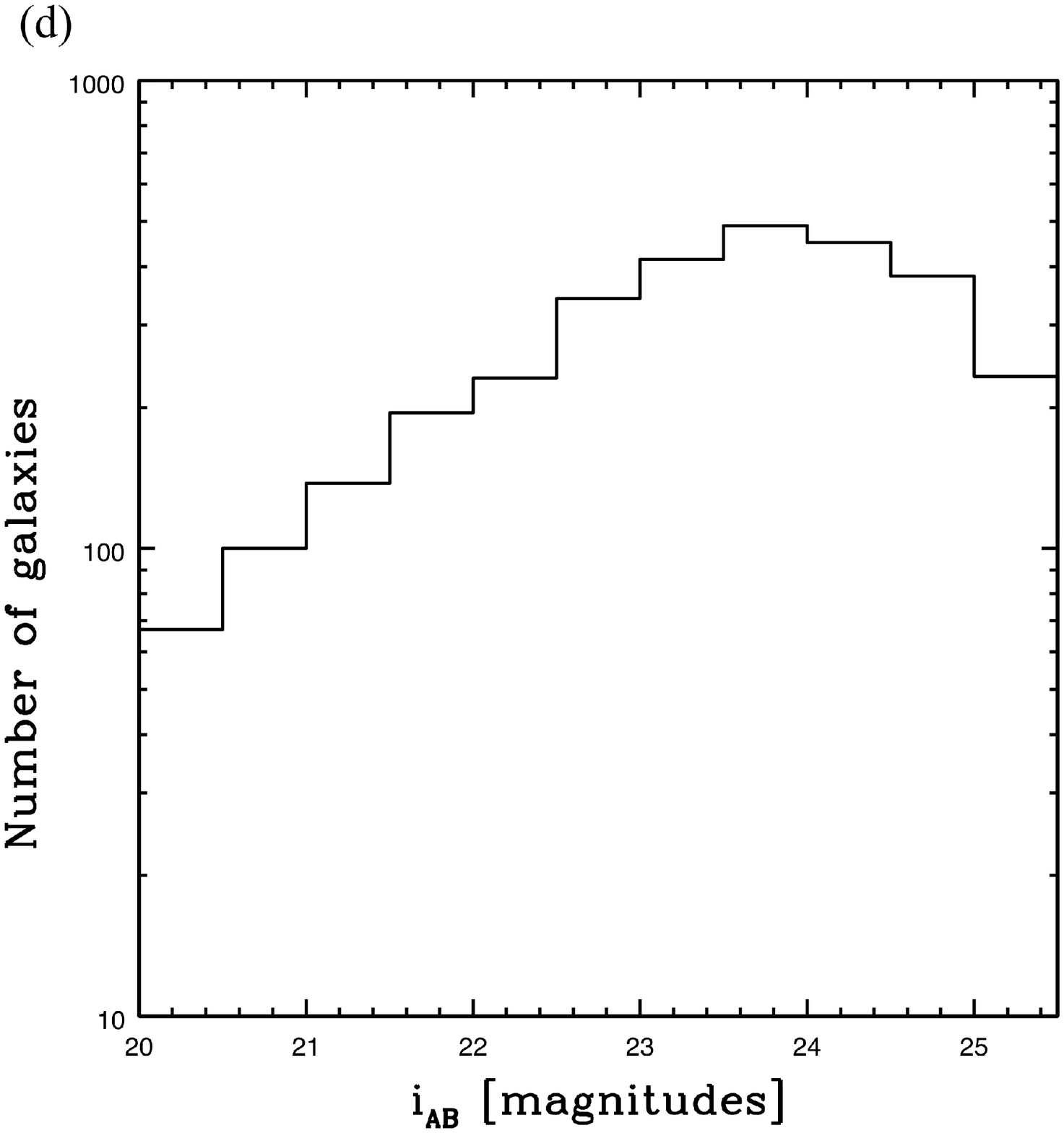}
\hspace{0.1cm}
\includegraphics[width=0.35\textwidth]{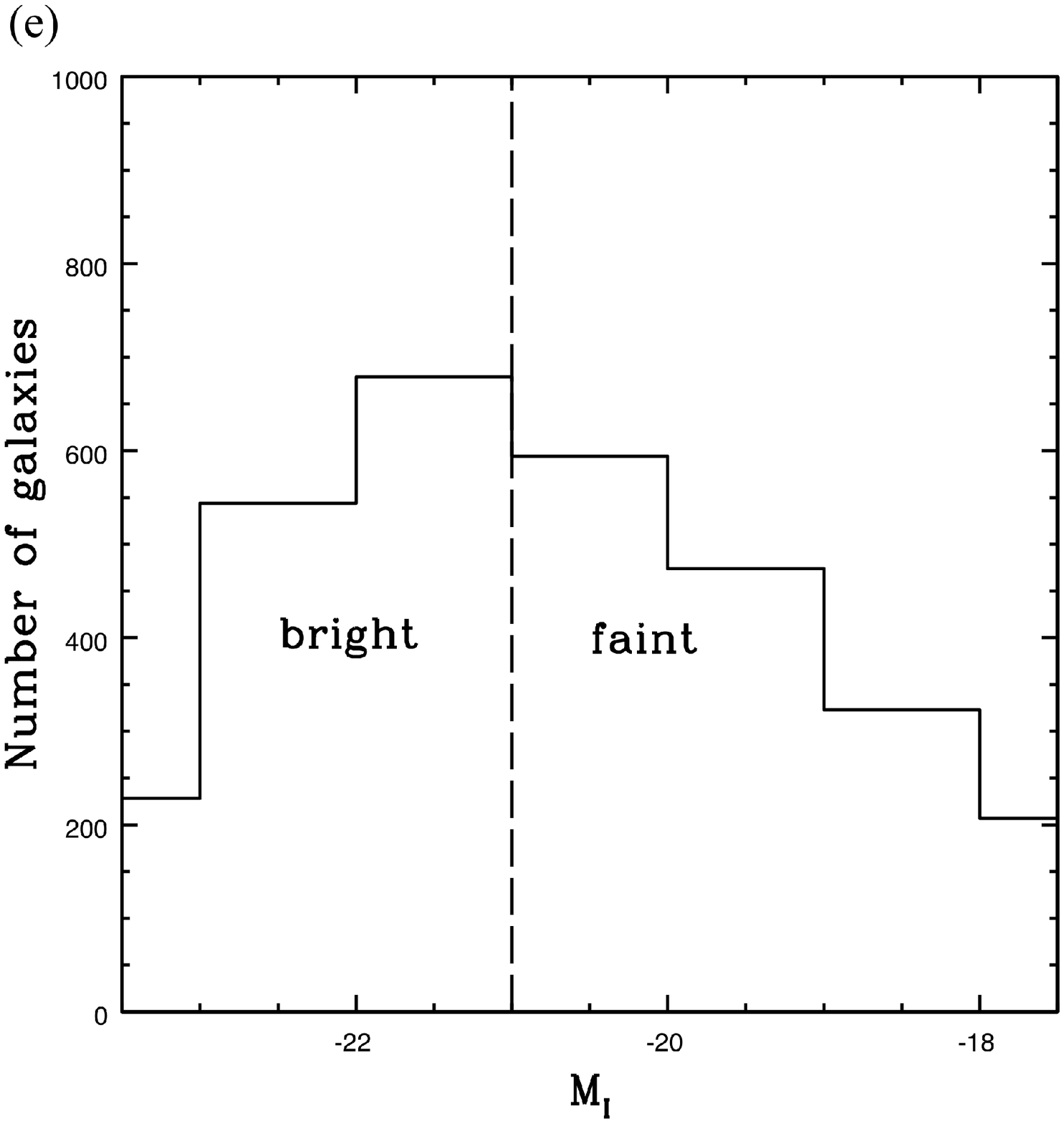}
\hspace{0.1cm}
\includegraphics[width=0.35\textwidth]{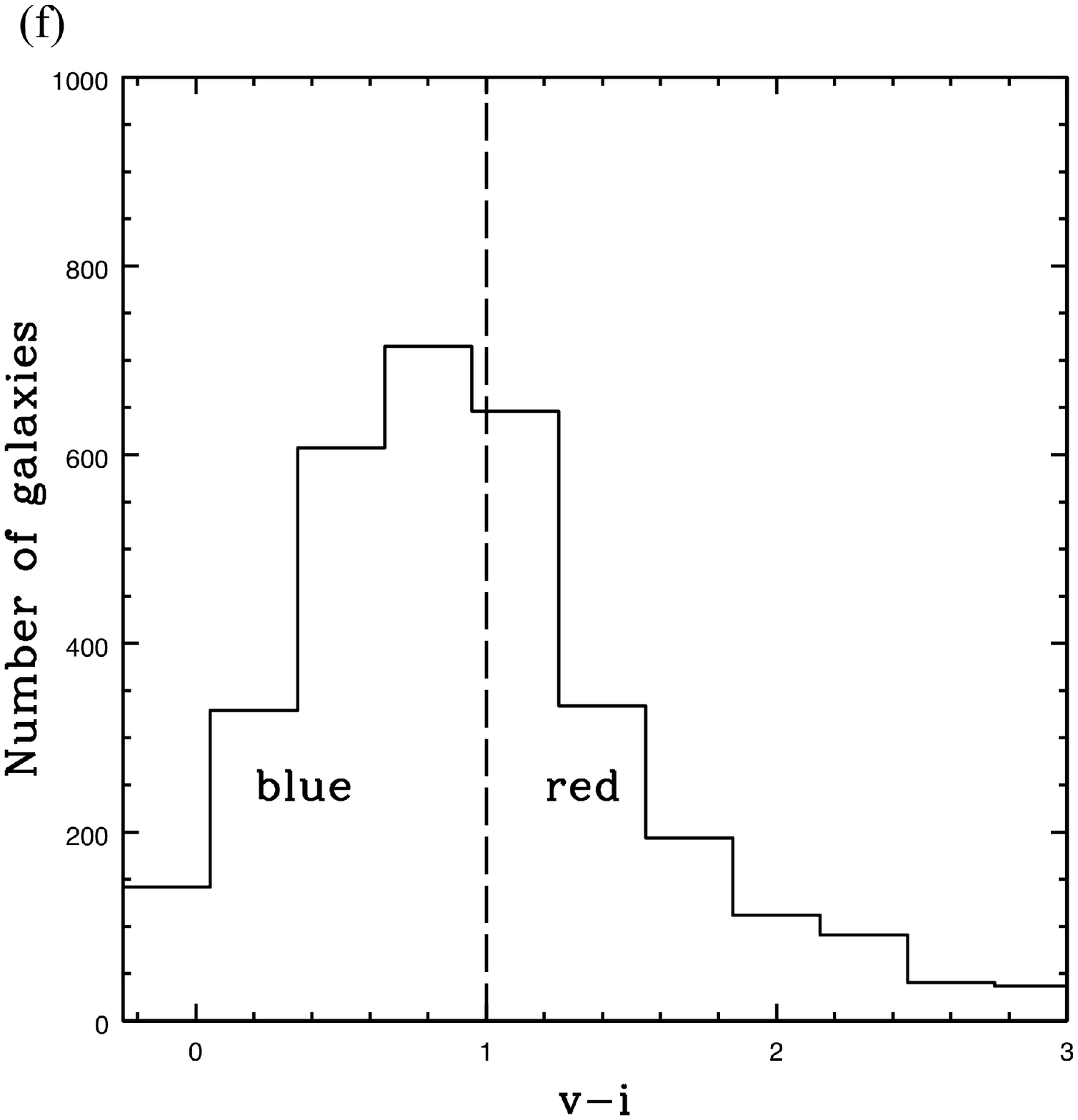}
}
}
\tiny{
\caption{
The distribution of the photometric properties of the galaxy sample in
this study. The panels show (a) the distribution of the Sersic index $n_{Sersic}$ (b) the
redshift distribution, showing the contribution of the photometric sample (solid
black line), the spectroscopic sample (dotted black line), the red
galaxies in the photometric sample (with $v-i\geq1.0$, red dashed line)
and the blue galaxies ($v-i < 1.0$, blue dashed line) (c) the
distribution of the half light radius $R_{50}$ (d) the $i_{AB}$
magnitude distribution (e) the $I$ band
absolute magnitude distribution (f) the distribution of $v-i$ color. 
}
 \label{fig:filters}}
\end{figure}

\begin{figure}
%\centerline{\rotatebox{0}{
%\includegraphics[height=0.4\textwidth]{diskgalaxy_R50_015_CGF.ps}
%\hspace{0.1cm}
%\includegraphics[height=0.4\textwidth]{diskgalaxy_R50_035_CGF.ps}
%}
%}
%\vspace{0.1cm}
\centerline{\rotatebox{0}{
\includegraphics[width=0.8\textwidth]{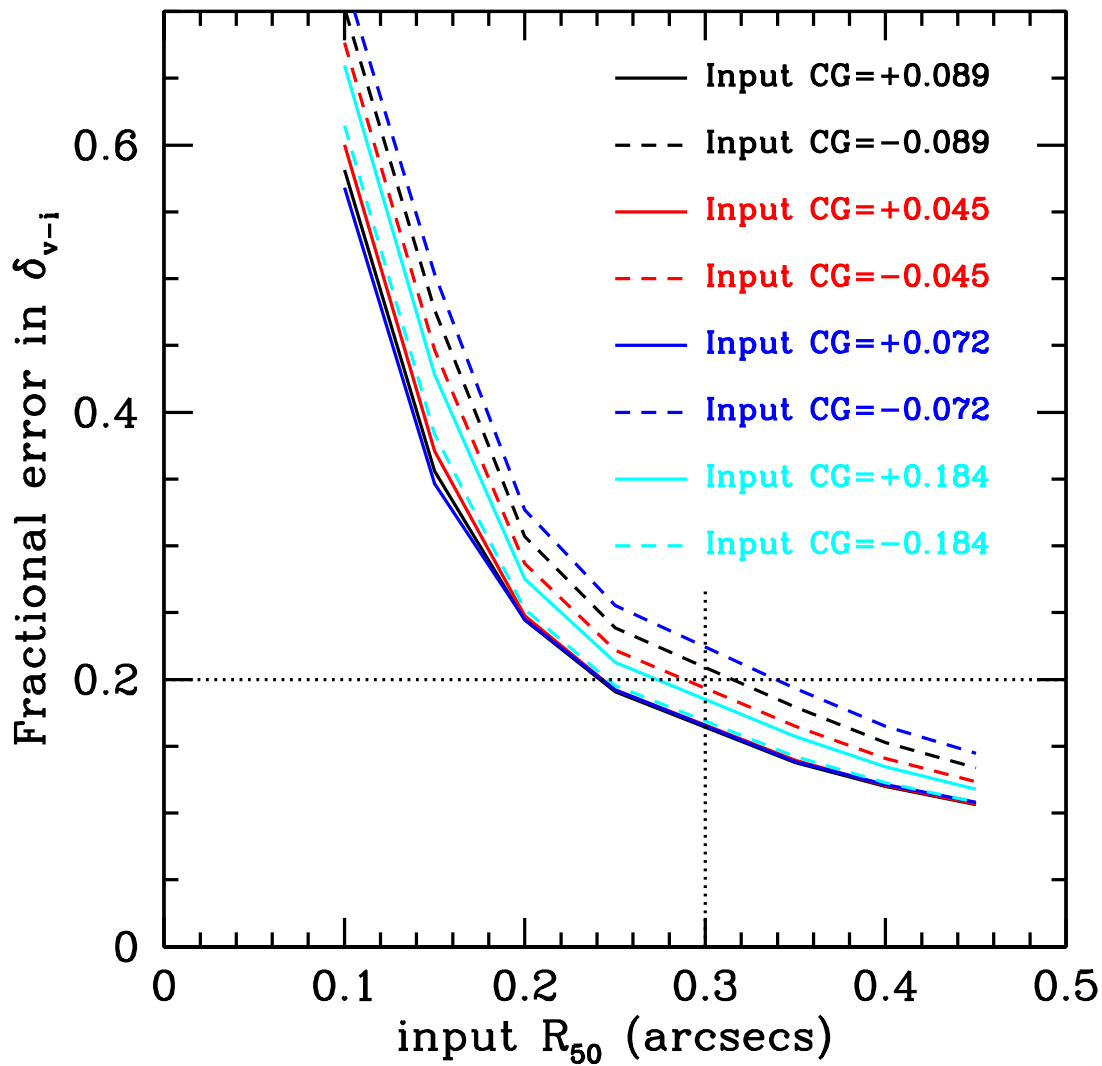}
\vspace{0.1cm}
}
}
\tiny{
\caption{Fractional error in the recovered $v-i$ color gradient as a
  function of the half light radius $R_{50}$ for simulated disk
  galaxies with an exponential radial surface brightness distribution. Marked is our
  selection in $R_{50}$ corresponding to a fractional error of 0.2 in
  recovering the input gradient (dotted line). 
%(b) the uncertainty in magnitude of the object in the $v$ and $i$ bands.
$v$ and $i$ band images of disk galaxies at $z=0.5$ have been simulated with various color gradients
assigned between $r<R_{50}$ and $R_{50}\leq r<2R_{50}$.
The images have the same pixel scale as the drizzled ACS $i775$ band
images (0.03 $\arcsec$/pixel). After assigning the input color
gradient, the image of the galaxy is smoothened by a Gaussian kernel
of $3.7\times3.7$ pixels in order to match the PSF FWHM in the $i775$ band ($\sim0.11\arcsec$). The luminosities in each pixel are derived
 from the Bruzual \& Charlot (BC03) stellar population synthesis
 models.  
}
\label{fig:simulations}}
\end{figure}

\begin{figure}
\centerline{\rotatebox{0}{
\includegraphics[height=0.8\textwidth]{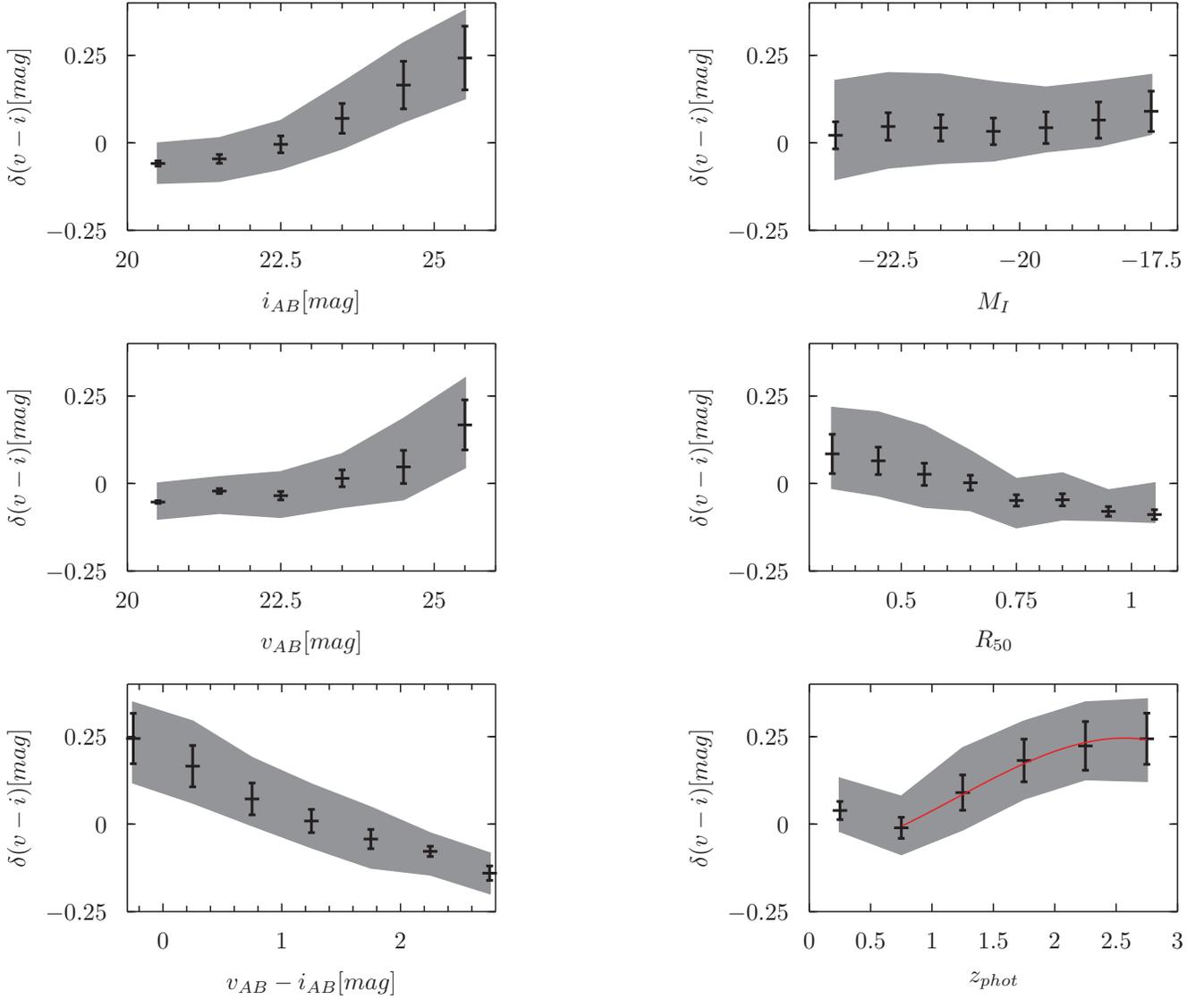}
}
}
\tiny{
\caption{Dependence of the color gradient $\delta(v-i)$ on galaxy properties in the
  GOODS-South photometric sample. The shaded region in each panel is
  bounded by the 25th and 75th percentile of the measured color
  gradient. The panels show the dependence of the gradient on (a) the
  $i_{AB}$ apparent magnitude (b) the $I$ band absolute
  magnitude (c) the $v_{AB}$ apparent magnitude (d) the galaxy half
  light radius $R_{50}$ (e) $v-i$ color (f) the photometric redshift $z_{phot}$. The redshift
  evolution of the median color gradient for $0.5<z\le3.0$ is described by the
  polynomial function $\delta_{(v-i)}(z) = -0.033z^3 + 0.119z^2 +
  0.045z - 0.093$ (red line). A bin size of $\triangle{z}=0.5$ is used
  and the data points correspond to the centres of the bins.}} 
\label{fig:color_gradient_trends}
\end{figure}

\begin{figure}
\centerline{\rotatebox{0}{
\includegraphics[height=0.32\textwidth]{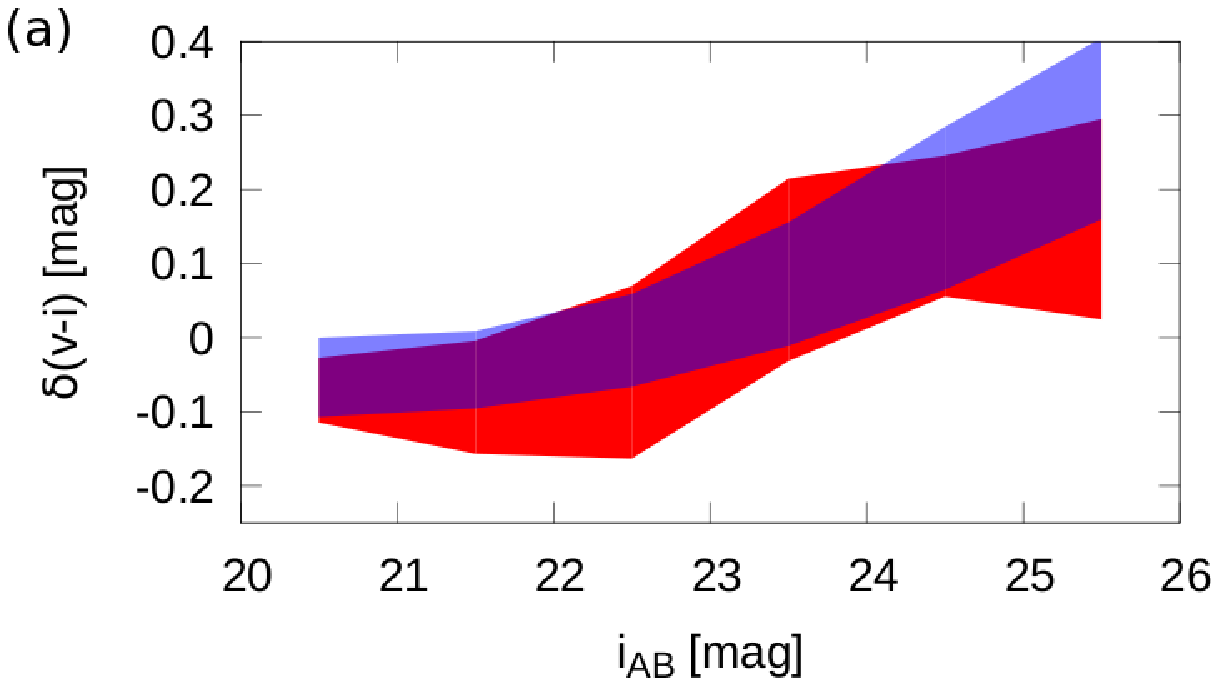}
\includegraphics[height=0.32\textwidth]{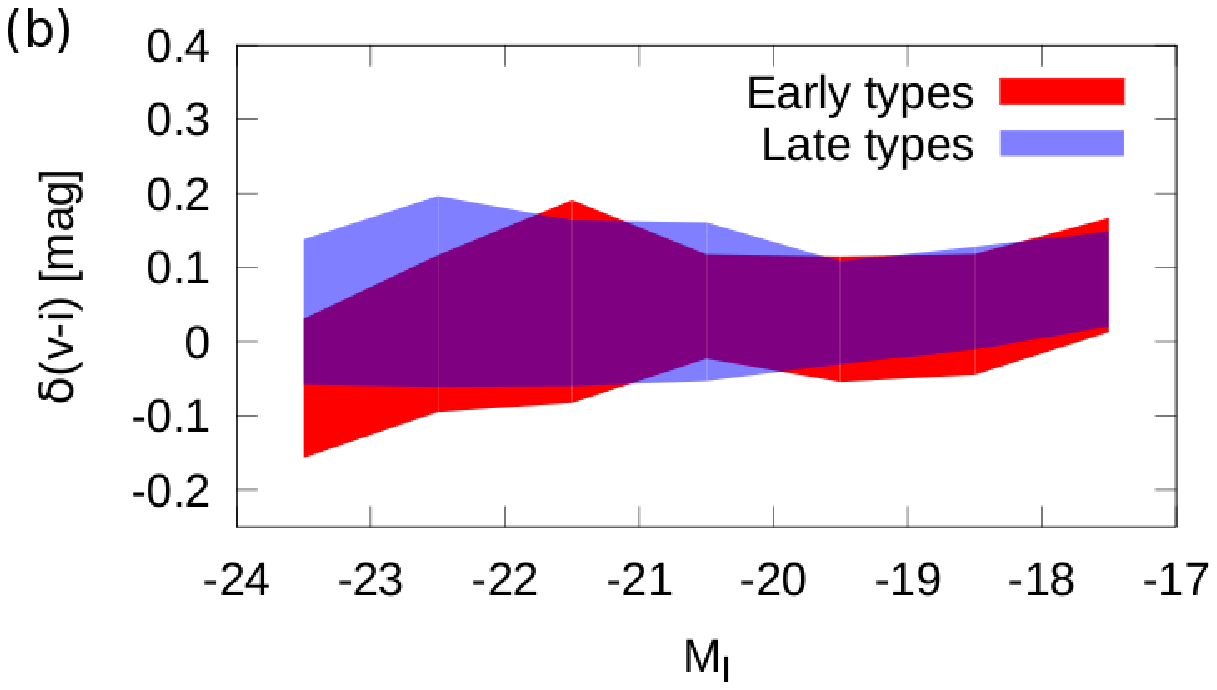}
}}
\vspace{0.1cm}
\centerline{\rotatebox{0}{
\includegraphics[height=0.32\textwidth]{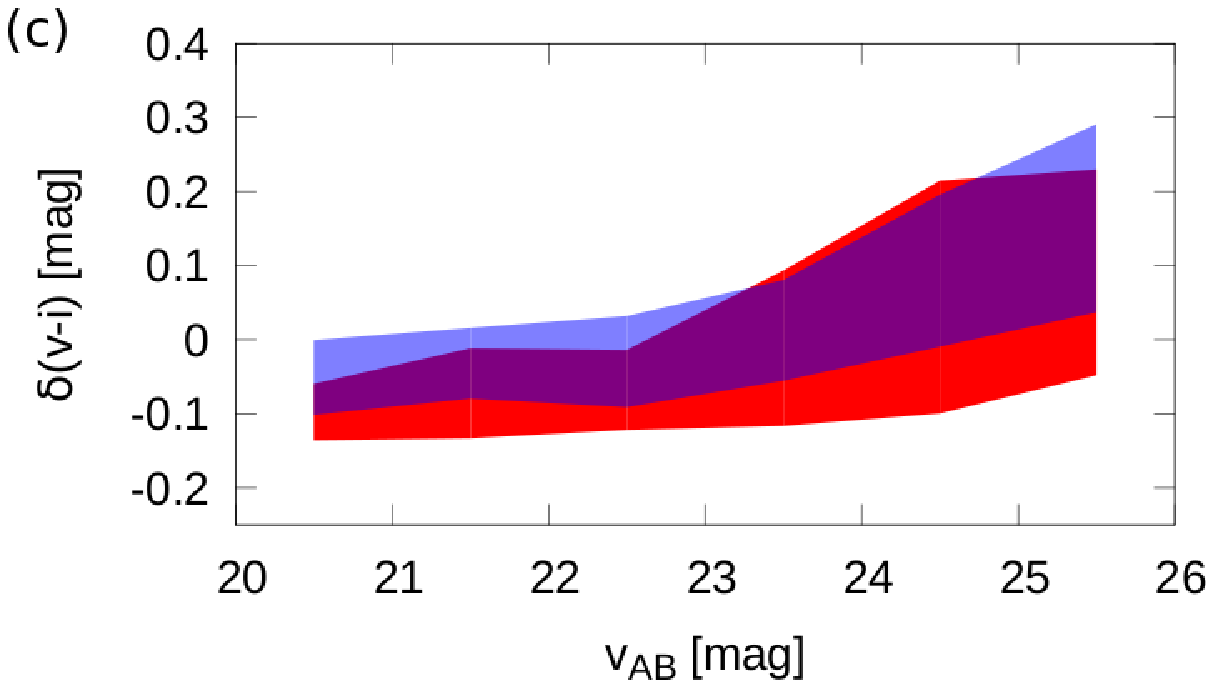}
\includegraphics[height=0.32\textwidth]{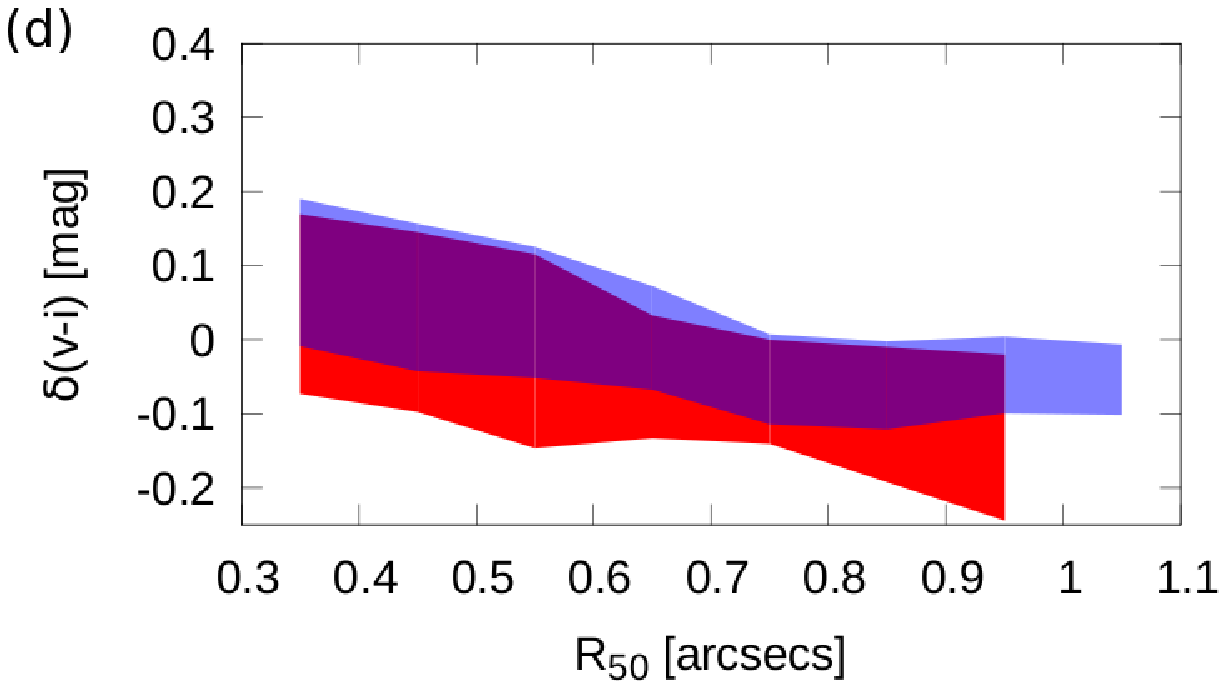}
}}
\vspace{0.1cm}
\centerline{\rotatebox{0}{
\includegraphics[height=0.32\textwidth]{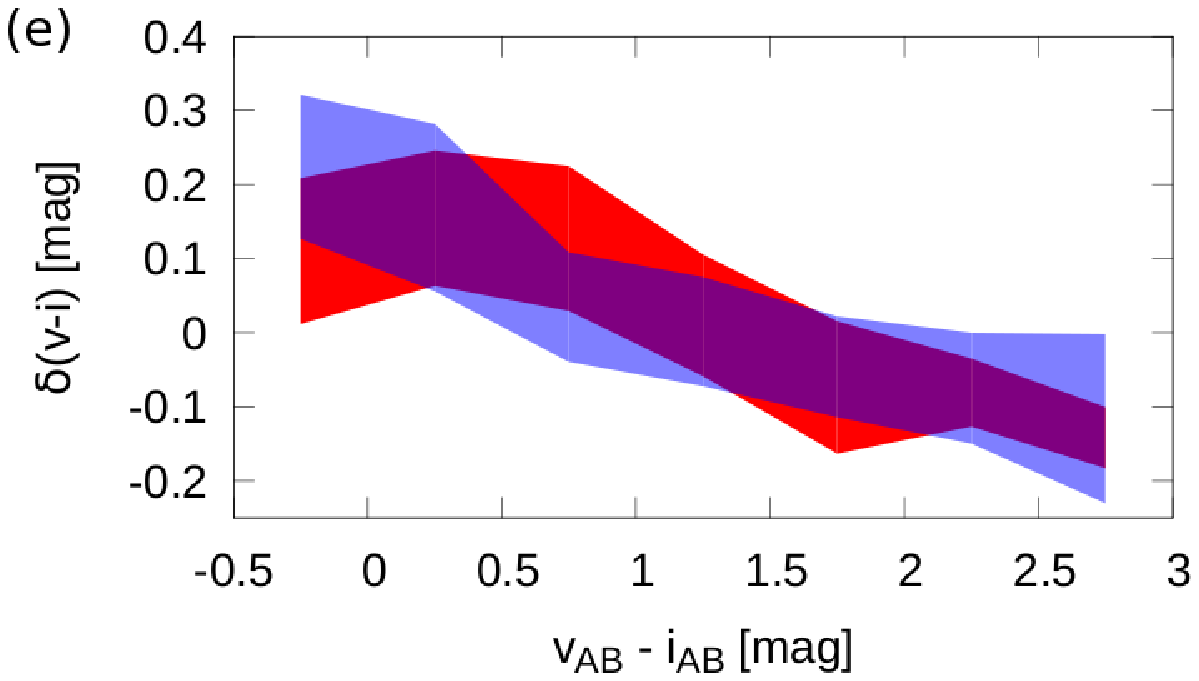}
\includegraphics[height=0.32\textwidth]{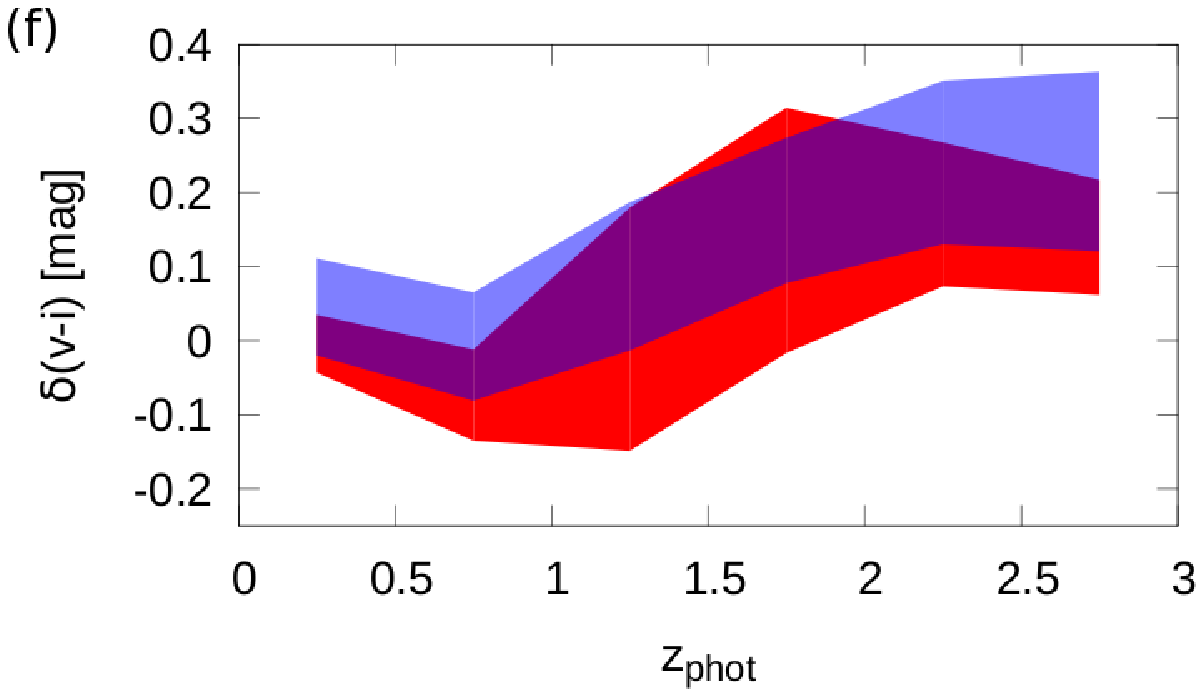}
}
}
\tiny{
\caption{Morphological type dependence of the trends in the color gradient. 
%(a) $i$ band magnitude (b) $v$ band
 % magnitude (c) $I$ band absolute magnitude (d) half light
  %radius (e)  $v-i$ color (f) $R_{50}$ (e) photometric redshift.
The panels show the dependence of the gradient on (a) the
  $i_{AB}$ apparent magnitude (b) the $I$ band absolute
  magnitude (c) the $v_{AB}$ apparent magnitude (d) the galaxy half
  light radius $R_{50}$ (e) $v-i$ color (f) the photometric redshift
  $z_{phot}$. The red shaded region corresponds to early-type galaxies in the
sample and the blue shaded region corresponds to late-type galaxies. 
The shaded region in each panel is bounded by the 25th and 75th
percentile of the color gradient in the sample being studied. 
}} 

\label{fig:color_gradient_trends_types}
\end{figure}

\begin{figure}
\centerline{\rotatebox{0}{
\includegraphics[height=0.55\textwidth]{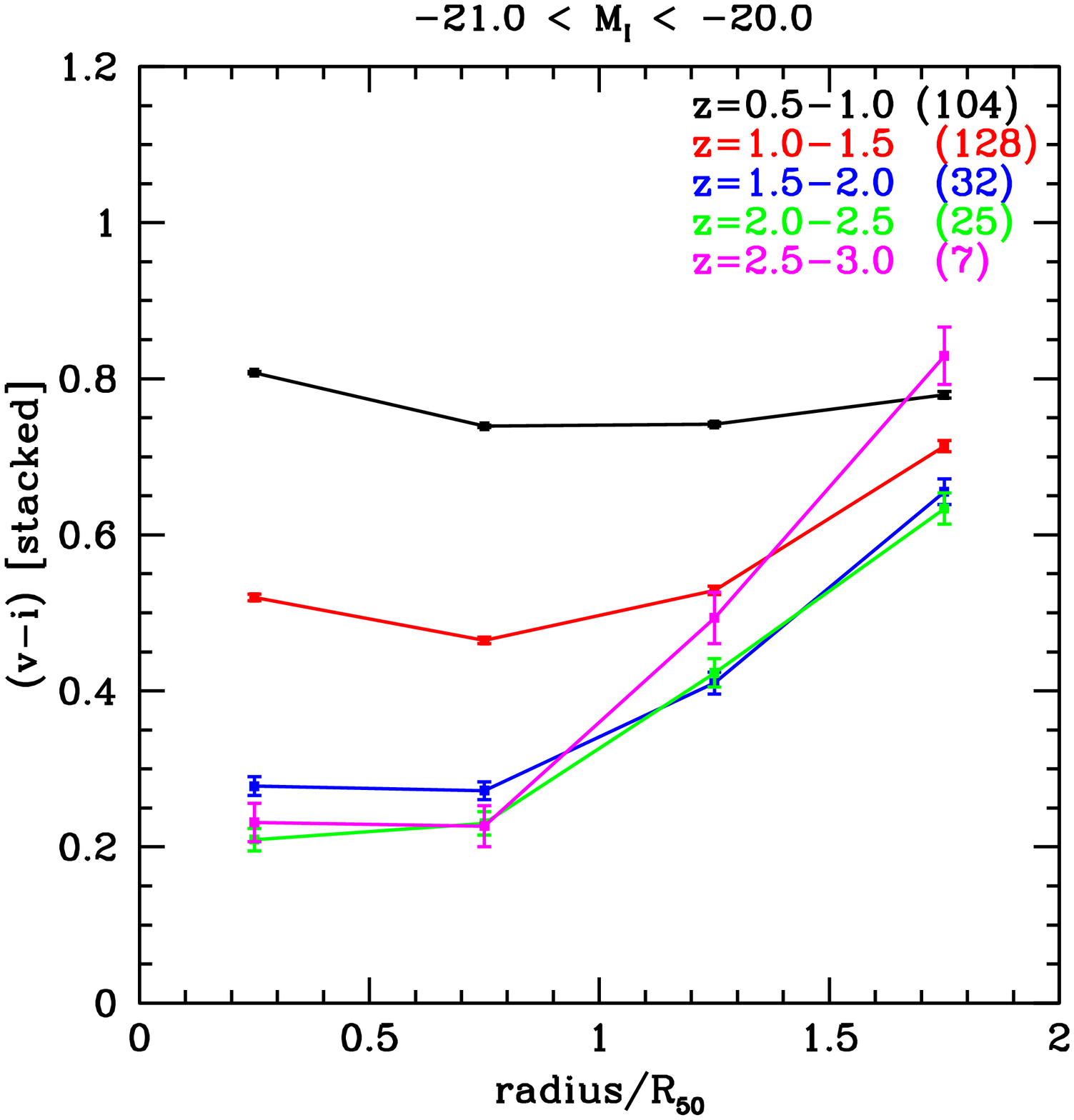}
}
}
\vspace{0.1cm}\centerline{\rotatebox{0}{
\includegraphics[height=0.55\textwidth]{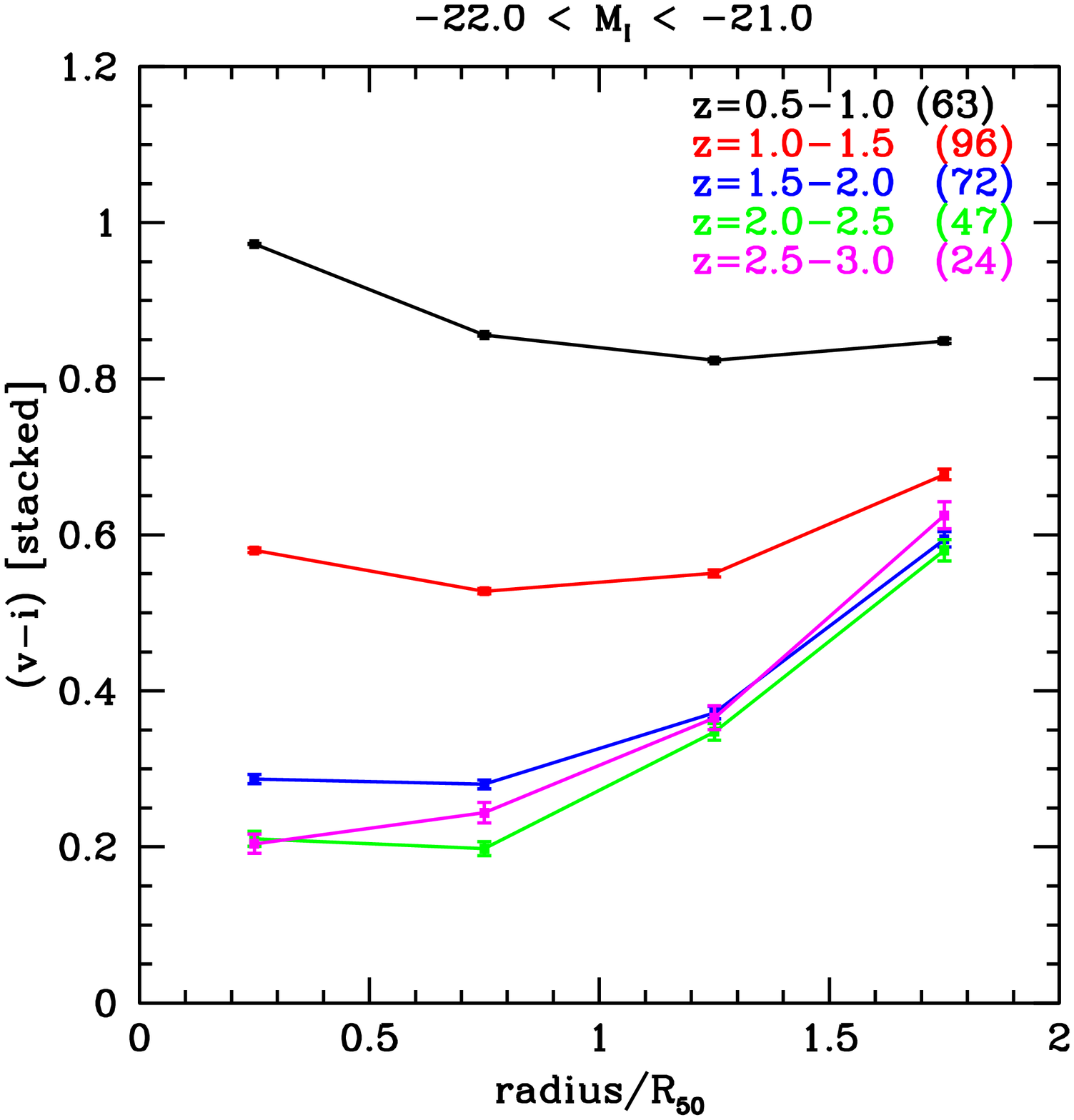}
}
}
\tiny{
\caption{Evolution of the radial variation of $v-i$ color in stacked galaxies
 with redshift for two different luminosity intervals. 
Galaxies are selected to have $R_{50}=0.30-0.40$ and to be in two
 intervals of absolute magnitude  $-21.0<M_{I} \le -20.0$ (faint bin, top panel) and $-22.0<M_{I}
  \le -21.0$ (bright bin, bottom panel). Galaxies in each of these
  absolute magnitude intervals have the radius of their annulli scaled by their $R_{50}$, before the 
  galaxies are stacked. The number of galaxies in each redshift bin
  is labelled in parentheses. At $z\sim2.0-3.0$, the SNR for the stacked
  galaxies in their outermost annuli ($r=1.5-2R_{50}$) is 9.0, and is above 2.0 for all radii up to $r=2R_{50}$.
}} 
\label{fig:color_gradient_trends}
\end{figure}

\begin{figure}
\centerline{\rotatebox{0}{
\includegraphics[width=0.8\textwidth]{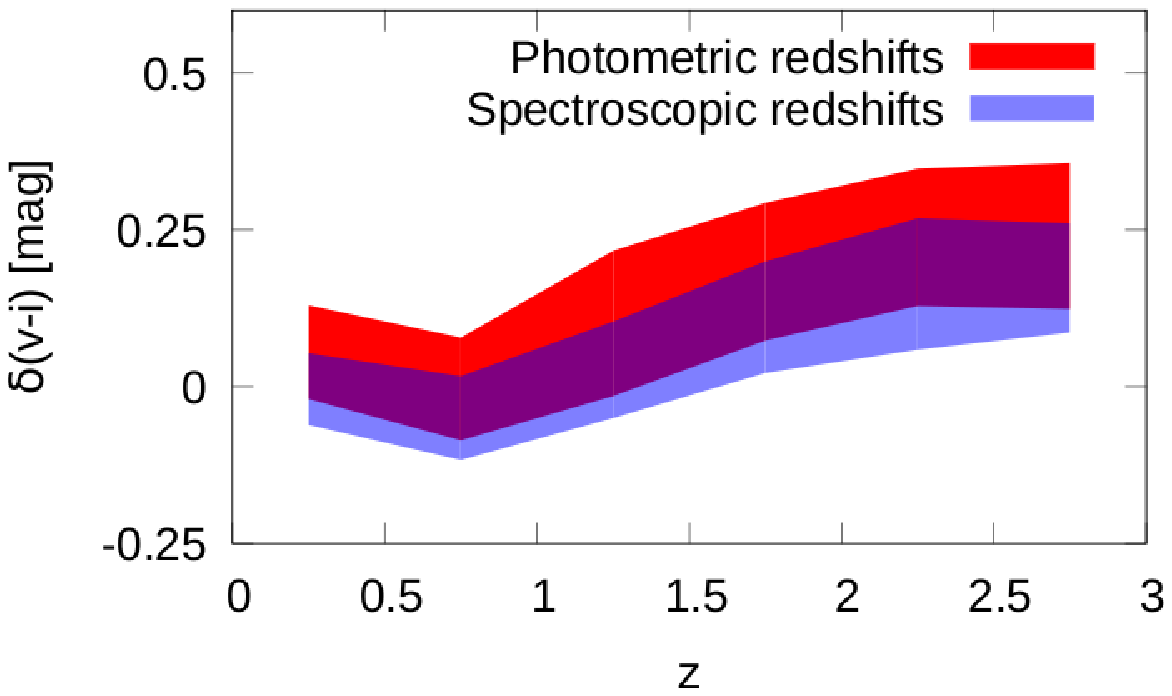}
}
}
\vspace{0.1cm}\centerline{\rotatebox{0}{
\includegraphics[width=0.8\textwidth]{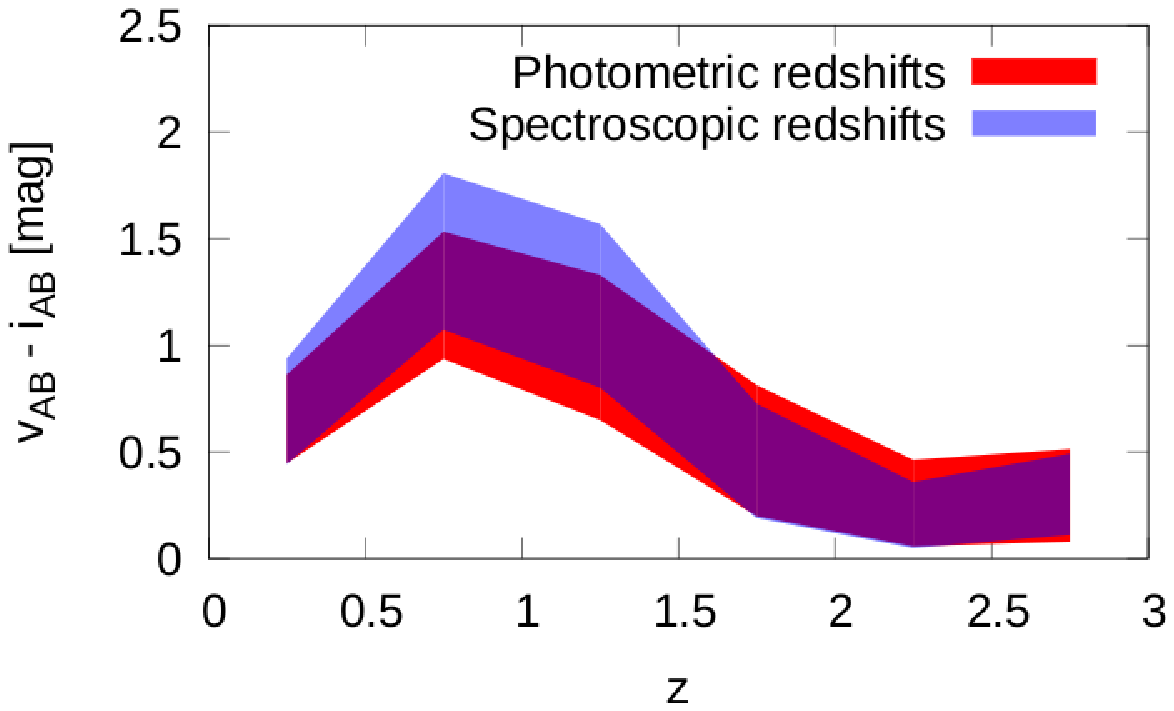}
}
}
\tiny{
\caption{Top panel: the color gradient $\delta(v-i)$ as a function of
  redshift when (a) photometric redshifts are used (red shaded
  region) and (b) only spectroscopic redshifts are used
  (blue shaded region). The spectroscopic redshifts have been derived from the GOODS-VIMOS spectroscopic campaign (Balestra et
  al.~2009). The bottom panel shows the color distribution $v-i$ of
  the sample as a function of the photometric redshift (red shaded region) and the
  spectroscopic redshift (blue shaded region).}} 
\label{fig:color_gradient_trends}
\end{figure}

\end{document}